\documentclass[roman,12pt]{article} 
\usepackage{titlesec}
\titlespacing\section{0pt}{10pt plus 4pt minus 2pt}{10pt plus 2pt minus 2pt}
\titlespacing\subsection{0pt}{5pt plus 4pt minus 2pt}{0pt plus 2pt minus 2pt}
\titlespacing\subsubsection{0pt}{5pt plus 4pt minus 2pt}{0pt plus 2pt minus 2pt}

\titleformat{\section}
  {\normalfont\fontsize{12}{15}\bfseries}{\thesection}{1em}{}
\titleformat{\subsection}
  {\normalfont\fontsize{12}{15}\itshape}{\thesubsection}{1em}{}
  \titleformat{\subsubsection}
  {\normalfont\fontsize{12}{15}\itshape}{\thesubsubsection}{1em}{}
\usepackage[english]{babel}
\usepackage[usenames,dvipsnames]{xcolor}
\usepackage[colorlinks=true,urlcolor=blue, citecolor=BlueViolet, linkcolor=blue]{hyperref}
\usepackage{dcolumn}
\usepackage{multirow}
\usepackage{color}
\usepackage{booktabs}
\usepackage{rotating}
\usepackage{epsfig}
\usepackage{appendix}
\usepackage{amsmath}
\usepackage{amssymb}
\usepackage{version}
\usepackage[square,longnamesfirst]{natbib}
\usepackage{titlesec}
\usepackage{amsfonts}
\usepackage{graphics}
\usepackage{fancyhdr,lastpage}
\usepackage{lscape}
\usepackage{bbold}
\usepackage{amsfonts}
\usepackage{color}
\usepackage{caption}
\usepackage{subcaption}

\usepackage{pdfpages}
\usepackage{etoolbox}
\AtBeginEnvironment{thebibliography}{\linespread{1}\selectfont}

\usepackage{pgf}
\usepackage{tikz}
\usepackage{setspace}
\usepackage{multirow}
\usepackage{longtable}
\usepackage{varioref}
\usepackage{lscape}
\usepackage{natbib}
\usepackage{amsmath, amssymb, bm}
\usepackage{rotating}
\usepackage{enumerate}
\usepackage{amsthm}
\usepackage{graphicx}
\usepackage{authblk}
\usepackage[T1]{fontenc}
\usepackage[utf8]{inputenc}
\usepackage{calc}
\usepackage{import}

\usepackage[margin=1in]{geometry}
\usepackage{todonotes}

\usepackage{booktabs}
\usepackage{rotating} 
\usepackage{lscape} 
\usepackage{longtable}

\usetikzlibrary{shapes,arrows,positioning,matrix,fit,automata}
\tikzset{
        >=stealth',
        punkt/.style={
           rectangle,
           rounded corners,
           draw=black, very thick,
           text width=6.5em,

text height=2em,
           text centered},
        pil/.style={
           ->,
           thick,
           shorten <=2pt,
           shorten >=2pt,}
}
\tikzstyle{sensor}=[draw, fill=blue!20, text width=5em,
    text centered, minimum height=2.5em]
\tikzstyle{ann} = [below, text width=5em]
\tikzstyle{naveqs} = [sensor, text width=5em, fill=red!20,
    minimum height=2.5em, rounded corners]
\tikzstyle{plain}=[draw,text width=5.1cm,
    text centered, minimum height=2.5em]

\newlength{\yellownotewidth}
\newlength{\yellownoteheight}
\usepackage{todonotes}

\bibpunct{(}{)}{;}{a}{,}{,}

\title{Editing a Woman's Voice\thanks{The authors thank Jiarui Liu, Kyle Lo, and Justus Mattern for their valuable research assistance. We appreciate comments from participants at the Conference on Empirical Methods in Natural Language Processing. We also acknowledge useful discussions with John Barrios, Mihir Mehta, Michael Minnis, Pietro Veronesi, and Luigi Zingales.}}
\author[1]{Anna M. Costello}
\author[2]{Ekaterina Fedorova}
\author[3]{Zhijing Jin}
\author[4]{Rada Mihalcea}

\affil[1,2]{University of Chicago, Booth School of Business}

\affil[3]{ETH Zurich, Department of Computer Science}
\affil[3]{Max Planck Institute for Intelligent Systems, Tubingen, Germany}
\affil[4]{University of Michigan, Department of Computer Science and Engineering}

\date{March 2023}

\begin{document}
\maketitle

\bigskip
\vspace*{-1cm}
\begin{abstract}
\begin{singlespace}
Prior work shows that men and women speak with different levels of confidence, though it's often assumed that these differences are innate or are learned in early childhood. Using academic publishing as a setting, we find that language differences across male and female authors are initially negligible: in first drafts of academic manuscripts, men and women write with similar levels of uncertainty. However, when we trace those early drafts to their published versions, a substantial gender gap in linguistic uncertainty arises. That is, women \textit{increase} their use of cautionary language through the publication process more than men. We show this increase in the linguistic gender gap varies substantially based on editor assignment. Specifically, our author-to-editor matched dataset allows us to estimate editor-specific fixed effects, capturing how specific editors impact the change in linguistic uncertainty for female authors relative to male authors (the editor's author-gender gap). Editors' author-gender gaps vary widely, and correlate with observable editor characteristics such as societal norms in their country-of-origin, their work history, and the year that they obtained their PhD. Overall, our study suggests that a woman's ``voice'' is partially shaped by external forces, and it highlights the critical role of editors in shaping how female academics communicate.
\end{singlespace}   
\end{abstract}
\bigskip
\bigskip

 \textbf{Keywords:}	Hedging; NLP; Peer Review; Bias

\thispagestyle{empty}

\setstretch{1.5}
\newpage
\setcounter{page}{1}

\section{Introduction}
Why do men and women speak differently? There is an extensive body of literature on the socialization of children into traditional male and female roles, which includes differences in linguistic traits. In her seminal study ``Language and a Woman's Place,'' \cite{lakoff:1973} states that ``In appropriate women's speech, strong expression of feeling is avoided, expression of uncertainty is favored, and means of expression in regard to subject-matter deemed trivial to the real world are elaborated.'' Yet, little is known about the extent to which females conform to or resist linguistic norms, especially in professional environments where they are taught to `lean in.'\footnote{See https://leanin.org/ and https://money.usnews.com/money/blogs/outside-voices-careers/articles/2018-06-12/c-suite-advice-to-women-in-the-workplace-dont-hedge-your-message.} Further, outside pressures -- such as expectations by managers about how employees should behave -- may cause women to act with more uncertainty relative to their male counterparts.

We examine gender-based linguistic differences in the context of academic publishing. Specifically, we ask whether female authors are more likely to hedge their inferences -- e.g., write with more caution -- relative to their male counterparts. Our headline finding is the following: in first drafts of academic manuscripts, male and female authors write with similar levels of uncertainty. However, when we trace those early drafts to their published versions, a significant gender-based difference in the use of uncertain language arises: female authors write with an average of 11 more uncertain terms than male authors in published manuscripts. What causes this differential, gender-based change in hedging language over the course of the review process? Our paper explores the potential role of the editor in charge of evaluating the manuscript in contributing to the linguistic wedge between men and women.

Our sample starts with articles published in the top fifteen economics, finance, accounting, and management journals for the period 2014-2019. We classify authors' genders using name-to-gender inference services. To establish a link between the editorial process and gendered writing styles, we need a picture of what the manuscript looked like \textit{before} it was submitted to the journal. Thus, we scrape various academic websites to match each published article to their earliest, working-paper version of the same study (henceforth, the `pre-submission' version). Finally, to assess the role of the editor in contributing to an author's cautionary language, we extract the editor's name from the acknowledgements and footnotes of each manuscript.

The specific linguistic feature we examine is textual uncertainty, which we capture in two ways. First, we measure the author's use of hedging language, a common linguistic practice that indicates uncertainty or cautionary language. Unlike a simple bag of words, detecting uncertainty through hedge words requires an analysis of the surrounding context. We use a dataset of annotated uncertainty cues and a fine-tuned SciBERT model to detect the degree of uncertain language for each article. BERT, or Bidirectional Encoder Representation from Transformers, is a pre-trained language model that learns the context of words bi-directionally. Second, we supplement the SciBERT model with a dictionary of 13 phrases commonly used in economics-based articles to explicitly caution or caveat a study's inferences. These additional words fall outside the concept of hedging, but are commonly used to dampen inferences (e.g., ``caveat''). For each full-text article -- both pre-submission and published versions --  we construct a summary measure of \textit{Uncertainty}, computed as the combination of hedge detection cues from the SciBERT model and our supplementary bag-of-words that capture caveats. 

As highlighted above, our descriptive analysis shows that men and women exhibit similar levels of uncertainty in their pre-submission versions, but then a large and statistically significant gender-based wedge in uncertain language emerges in the published versions of those same articles. It's important to recognize that changes in uncertainty may be explained by differences in the fundamental aspects of the project, such as the topic, the method, and the quality of execution. Indeed, prior work highlights the important role of hedging in correctly framing the inferences from scientific studies \citep{hyland1996a,hyland1996b,hyland1998,salager1994,varttala2001}. Should men and women write fundamentally different types of papers, the differential changes in cautionary tone may be expected. However, an \textit{overly} cautious tone -- e.g., one that is not warranted by the fundamental aspects of the study -- can dampen the study's merits and reduce the author's contribution to the scientific community. We explore whether the differential change in cautionary language for females relative to males is at least partially due to factors unrelated to fundamental aspects of the study. 

We start by explaining changes in uncertainty in a regression framework that controls for a multitude of study characteristics, including the topic, the year it was published, the journal it was published in, the time that the paper spent in the review process, and the quality of the author team (measured by citations). After controlling for these characteristics, we continue to find that female authors increase their level of written uncertainty more than male authors.\footnote{We also use a specification that controls for author fixed effects. That is, we expand the data so that each observation represents a version of the manuscript (pre-submission and published) rather than each observation being the change in a single manuscript. We then run a difference-in-differences specification that interacts an indicator variable for a female author with a ``published'' indicator to capture the change in uncertain words for the published version relative to the pre-submission version. Because each author has at least two versions of a manuscript, we are able to include an author-level fixed effect that controls for the average, within-author writing style, quality, etc.}  They do so to a marginal degree in the abstract, but more so in the main text, and specifically in the conclusion section of the manuscript. 

We further recognize that, though comprehensive, our control variables may not fully subsume the fundamental, expected variation in changes in uncertainty of academic articles.\footnote{For instance, our control variables only noisily capture differences in topic, approach, and quality.} Our ideal setup would compare a male-written manuscript to a counterfactual female-written manuscript that is equivalent along all dimensions except for the author's gender. We approach this challenge using natural language processing tools that can more robustly capture the ``fundamentals'' of an article and thus more precisely and separately isolate the causal effect of gender on changes in uncertainty. Specifically, we use a SciBERT model to create a numerical representation of the text in a 768-dimensional vector space -- capturing features like vocabulary, writing style, research approach, and quality (henceforth, the text-representation vector). We supplement the text-representation vector with other control variables like author citations, time spent in review, etc. Together, these variables more fully subsume the fundamental aspect of the paper, and any remaining explanatory power of the author's gender on changes in uncertainty should be unrelated to fundamentals. We then train a model using deep learning to predict the change in uncertain words for an article -- using both fundamentals and the author's gender as inputs.\footnote{Our model achieves 81\% accuracy in predicting uncertainty words.} In a second step, we use the trained model to predict the number of uncertainty words for a male-authored paper, should that paper be authored by a female. Specifically, we hold the text representation vector and other fundamental control variables constant, artificially `flip' the gender of male-authored papers to be female-authored, and allow the model to predict the number of uncertain words for these pseudo-female observations.  This approach -- which compares the uncertainty of a male-authored paper to a nearly identical counterfactual article written by a hypothetical female -- yields similar results to our main regression analyses. It indicates that gender itself plays an important role in how articles change over time.

Next, we move to understanding the forces that cause females to increase their written uncertainty more during the editorial process, relative to male authors. In a wide range of studies, authors have documented the biases that female academics face during the pre-submission and editorial decision stages \citep{dupas2021,hengel2022,card2020,sarsons2017}. For instance, \cite{siniscalchi2021} show that evaluators' self-image bias can impact scholars' outcomes. Thus, there may be an important role played by the evaluator (e.g., editor) in promoting a larger increase in female cautionary tone. Alternatively, editorial feedback may be unbiased, and the changes in female exposition may be explained by other forces\footnote{For example, females may receive different feedback outside of the editorial process (\cite{dupas2021})}; this would suggest that the identity of specific editors -- and in particular his or her influence on authors' papers -- should not relate to females' larger increases in uncertainty. 

For each manuscript, we identify the editor in charge of evaluating the paper for publication. Intuitively, we want to identify the role of specific editors in contributing to the larger increase in uncertain language for female authors, relative to male authors. Our data, whereby we have editors that oversee multiple papers written by different authors, and authors with papers under review by different editors, is particularly well-suited to estimate an `editor effect.' Take a hypothetical editor named John: John accepts multiple papers, some written by males and some written by females. We can hold John's \textit{average effect} on the change in authors' writing constant (e.g., include a `John' fixed effect), and estimate John's differential influence on female-authored papers versus male-authored papers. At the same time, consider a hypothetical author, Sally. Sally submits multiple papers to journals and has some of them accepted, all by different editors. We can hold Sally's specific writing style, approach, and quality constant by including a `Sally' fixed effect. Under this fixed effect structure, we can then see how John's impact on an author's linguistic style varies when he is paired with Sally as an author, versus when he is paired with a different author in the dataset, all while holding John and Sally's average styles constant. From this specification, we extract an editor-specific estimate of their differential effect on female changes in uncertainty, relative to their effect on male changes in uncertainty (an \textit{editor x female author} fixed effect). Hereafter, we refer to this fixed effect as the editor's author-gender gap.

We show that the editor's author-gender gap explains a large portion of the variation in changes in uncertainty. Specifically, including an \textit{editor x female author} fixed effect in a determinants model explaining changes in articles' uncertainty words increases the adjusted R-squared of the regression by three percentage points. Further, the difference between male- and female-author changes in textual uncertainty varies considerably, depending who the editor of the paper is. An editor in the 75th percentile of the gender gap differentially increases female authors' uncertainty by 53 phrases, or 26\%, relative to an editor in the 25th percentile of the gender gap.

Because we also include author fixed effects and editor fixed effects in the regression, the editor-specific gender gap should isolate the impact of a particular relationship -- e.g., John matched with Sally -- on changes in writing style. It is worth noting that papers are not typically \textit{randomly} assigned to editors, but rather are assigned based on a match between the topic and the expertise of the editor. Endogenous matching may impact our inferences if the matching characteristics systematically correlate with fundamental differences across authors' genders. For example, assume authors have both high quality papers and low quality papers; inferences may be threatened if, e.g., John receives females' (lower quality and/or riskier) papers and males' (higher quality and/or safer) papers.\footnote{Endogenous matching concerns would have to satisfy the following conditions: (1) there is heterogeneity in the author's portfolio, (2) there is heterogeneity in the editor's expertise, and (3) the characteristics that create matches between authors and editors systematically line up with fundamental differences across the authors' genders.}

While we control for the paper's topic, the journal it was published in, and other fundamental characteristics, the endogenous matching concern may not fully resolve. Thus, the ideal empirical setting would involve random assignment of papers to editors, ensuring the assignment of fundamentally similar male and female papers to each editor. We sent an email to the managing editor of each journal to inquire about how they assign papers to editors. Two journals responded that assignment was purely random. For these two journals that randomly assign manuscripts to editors, we test whether these editors exhibit significant author-gender gaps; indeed, we continue to find that the \textit{editor x female author} fixed effect explains a large and statistically significant portion of the variation in changes in linguistic uncertainty, suggesting that endogenous matching does not fully explain our results.

Having established that editors themselves contribute to differential changes in female authors' uncertainty, we next move to exploring whether the gender gap lines up with specific editor types. In other words, is there something about an editor's background that makes him or her more likely to penalize female authors? We explore four such characteristics: (1) the gender equality index from the editor's country of origin; (2) the gender of the editor; (3) the editor's work history, specifically their pattern of coauthorships; and (4) the year the editor graduated with their PhD. We indeed find that the gender gap is correlated with traits of the editor. In particular, editors from low gender-equality countries impose a larger gender gap, relative to editors from high gender-equality countries.  Editors' past work experience -- specifically those that have a history of selecting primarily male coauthors -- also impose a larger gender gap. Finally, editors that graduated in more recent years have a lower gender gap.

The results are open to several alternative interpretations. Of note, while we try to attribute the gender gap to the actions of the editor, it's important to recognize that the exposition of a manuscript is both a function of editorial input and an author's response to that input. One may be concerned that our specific editor effects may still reflect authors' responses rather than biased actions on the part of the editor. For instance, assume that certain editors are more critical -- to both female and male authors -- than other editors; if female authors are more sensitive to criticism than male authors, this may result in an author-gender gap that is larger for those more critical editors.  We provide preliminary evidence that this alternative story does not explain our result. In particular, the gender gap does not correlate with proxies for an editor's harshness. A second possible interpretation of our results is that editors with a larger gender gap impose a lower acceptance bar for female-authored papers, but then they appropriately require more cautious language. If this were the case, we would expect editors with higher gender gaps to also have higher female acceptance rates; however, our data does not support this alternative story.

Our paper contributes to three strands of literature. First, we contribute to the literature that explores gender-based linguistic differences. There is some disagreement in the literature on whether females speak with more uncertainty relative to males, with several studies providing affirmative evidence (e.g., \cite{mirzapour2016}; \cite{rosanti2016}) and others finding a null result (\cite{leaper2011}). We add to this literature by showing that context matters: female academics do not \textit{initially} differ from male academics in their use of cautionary language. The review process itself, and specifically the guidance from specific editors, results in more hesitant language used by female authors. This finding is important because it is commonly believed that \textit{inherent} female traits -- e.g., risk aversion or a gender confidence gap -- drive their hesitancy, a phenomenon known as internal silencing (\cite{murciano2022}; \cite{exley2022}). However, our work points to a different underlying mechanism: the difference between male and female certainty is in part due to external influence imposed by editors. 

Second, we contribute to the literature on gender bias. Female academics encounter biases in the form of lower acceptance rates, less credit for group work, and harsher feedback during presentations (e.g., \cite{card2020}; \cite{sarsons2017}; \cite{dupas2021}). We explore a specific type of external bias that places limits on how females can frame their study, relative to how males can frame theirs. Thus, our project provides complementary evidence to prior work showing that females get less airtime in conference and seminar settings, they are more likely to be interrupted, and they face backlash if they are perceived as too assertive (\cite{jacobi2017}; \cite{davenport2014}; \cite{hinsley2017}; \cite{moss2012}). 

Third, we contribute to the literature documenting that managers have specific ``styles'' which impact the way corporations are operated and how they perform (\cite{bertrand2003}). As gatekeepers to the academy, editors make critical decisions that impact who gets hired and promoted through the tenure process, and thus they ultimately shape the composition of our academic institutions. Yet, we are unaware of any prior study that investigates the direct role of individual editors in shaping authors' outcomes. The granularity of our data allows us to estimate an editor-specific fixed effect that documents the direct effect that an editor has on an author's manuscript. Our results indicate that some editors have large author-gender gaps, and that this behavior is correlated with the editor's background and characteristics. In light of our results, we advocate for greater oversight and transparency into the editorial process.

The paper proceeds as follows. In section \ref{sec: framework} we discuss the framework linking gender and language patterns, in section \ref{sec: data} we present the data, in section \ref{sec: method} we outline our empirical methodology, and in section \ref{sec: results} we present our baseline results. In section \ref{sec: editor} we present arguments for and evidence of the role of the editor in gender-based linguistic differences, in section \ref{sec: robust} we outline several robustness tests, and in section \ref{sec: concl} we conclude and discuss our main inferences.
\section{Gender and Language \label{sec: framework}}

Gender differences in language patterns have been extensively studied in the psychology and linguistics literatures, starting with an early seminal study by \cite{lakoff:1973}. She proposed that gendered linguistic differences can be explained by roles and expectations in society, whereby men communicate in an assertive manner because they occupy the dominant position in the social hierarchy. She highlights that women use expressions of uncertainty and hedges in order to downplay their authority or to avoid being perceived as overly aggressive. Her hypothesis is that females are socialized at a young age, and that these language patterns persist into adulthood such that males speak with dominance and females speak more submissively.

Empirical tests of Lakoff's hypothesis has been mostly limited to small case studies and evidence is mixed, with some finding that females use more tentative language than males, and others finding negligible linguistic differences. Further, very few studies explore linguistic differences in professional settings where females and males hold relatively similar levels of education and other qualifications. The goal of our paper is twofold: first, we formally test for differences in language patterns across gender in a large sample of academic writing, with the goal of robustly isolating the role of gender in uncertainty. Second, we seek to explore the role that \textit{external parties} -- specifically editors -- play in contributing to variation in hesitant language. 

Related to the latter goal of our study, some evidence indeed suggests that societal pressures to conform to language norms persist into adulthood. For instance, small scale studies and anecdotes illustrate that men in the workplace are often praised for assertiveness, taking more ``airtime,'' and interrupting, whereas female leaders are punished for the same behavior -- known as the assertiveness double-bind (e.g.,\cite{brescoll2011}; \cite{amanatullah2013}; \cite{caprino2017}).

\subsection{Measuring Uncertainty}

Our key construct aims to reflect the concept of tentative language referred to by \cite{lakoff:1973}. We rely on two methods to detect uncertain language; the first measures the use of hedge words and phrases, and the second supplements hedges with a bag of words.

\subsubsection{Hedging}
We explore the concept of hedging, which has long been studied in the computational linguistics literature as a way to distinguish between fact-based and uncertain statements. Uncertainty through hedges can take many forms such as auxiliaries (e.g., may, might), verbs (e.g., suspect, appear, indicate), adjectives or adverbs (e.g., possible, unsure), conjunctions (e.g., or, either...or) or phrases (e.g., raises the question of). 

Our key challenge is to translate large volumes of text into a numerical representation such that we can measure variation in the use of hedge language. To do so, we rely on natural language processing (NLP) tools and specifically, a SciBERT model. BERT, or Bidirectional Encoder Representations from Transformers, ``learns'' the context of words (also referred to as tokens) bi-directionally. Specifically, BERT uses a deep neural network with the transformer architecture (\cite{vaswani2017attention}), which takes as input all the text tokens and applies 12 layers of fully connected attention mechanisms. The depth and connectedness of the network enables the model to capture complicated and long-range dependencies in text. 

A BERT model is ideal in our setting, because the context of words can change whether they should be categorized as hedging or not. Take the following example:
\begin{enumerate}
\item{You \textit{may} leave.}
\item{This \textit{may} indicate that.}
\end{enumerate}

The use of ``may'' only indicates uncertainty in the context of the latter statement, highlighting the importance of taking surrounding text into consideration in hedging classification. The SciBERT model we use leverages unsupervised pretraining on a large multi-domain corpus of scientific publications from Semantic Scholar (\cite{beltagy2019}). The model is then fine-tuned using a large dataset of text that was manually annotated to specifically flag uncertainty cues (\cite{zhizhin:2020}).\footnote{The annotation task was initiated as the Conference on Computational Natural Lanuage Learning (CoNLL) shared task in 2010. Text from WikiWeasel, Bioscope, FactBank, and other sources were annotated to tag uncertainty cues. See https://aclanthology.org/W10-3001.pdf. The dataset was later reannotated using additional data \cite{Szarvas2012}.} A visual overview of SciBERT model implementation can be viewed in Figure \ref{fig:intro_scibert}. The fine-tuned model is then used on our sample of academic articles for our downstream task of hedge detection. The hedge detection model takes a text input and returns the word(s) that function as uncertainty cues within that sentence, allowing us to construct a numerical representation of uncertainty for each article. 

\subsubsection{Bag of Words}
The hedge detection model was trained on domains adjacent to the field of academic economics. Specifically, the base SciBERT model was trained on publications from Semantic Scholar\footnote{Semantic Scholar consists of papers in the computer science field (approximately 18\% of the sample) and biomedical papers (82\% of the sample).}, and it was then fine-tuned on publications from Bioscope and Wikiweasel. Therefore, the hedging model may not fully capture the linguistic norms used by economists to indicate uncertainty. For instance, articles in the field of economics often include specific types of phrases to dampen inferences (e.g., \textit{caution, caveat}), which fall outside of the concept of hedging as employed in the fine-tuned SciBERT model. 

To ensure that we construct a measure of uncertainty that applies well to our sample, we randomly chose 100 articles to manually read and tag for uncertainty. We hired annotators to read the articles and tag any words or phrases that indicate an uncertain or cautionary tone. Each article had at least two annotators assigned.\footnote{In cases of disagreement, the authors made the classification decision.} We then collected a list of the most commonly used phrases in our sample that indicated uncertainty, but that were not picked up by our hedging model.\footnote{The annotation task produced the following list of additional tag words: caution, cautious, caveat, limit, confound, threat, concern, flaw, concede, acknowledge, aware, not generalize, cannot.} Words in the dictionary were detected in sentence inputs by using NLTK word tokenization and bigrams methods. Specifically, any instance of a word token or bigram that matched or contained a word in the dictionary was counted as a dictionary hit.

\section{Data \label{sec: data}}

We measure linguistic uncertainty in the fields of academic economics, finance, accounting, and management. To ensure that our sample comprises articles of similar quality and to make the scale of our study manageable, we select 15 top journals across these fields. Our goal is to create a dataset of presubmission-to-published article pairs, so that we can explore how uncertainty within a given article changes through the publication process. To do so, we start by downloading all published articles in each of the 15 journals for the five year period spanning 2015-2019. We then use Google Scholar to find the earliest (pre-submission) version of each published article. It is important to note that for the fields of economics, finance, accounting, and management, it is common for authors to post early, unpublished versions of their manuscripts online. In these fields, authors present unpublished work at conferences, other universities within their peer network, and in other public venues; those venues then host a version of the author's unpublished manuscript on their website.\footnote{Additionally, authors may post an early working paper on their website or on a third party host like NBER or SSRN.} This practice of publicizing early work makes our setting compelling from a data perspective, in that it allows us to gather and track an article's changes over time.\footnote{This contrasts with conventional practices in most other academic fields, whereby authors present only published studies, and/or don't post pre-submission studies.}

To collect the matched presubmission version of each published article, we build a web scraper to search Google Scholar for the article title, taking the first hit as the best match result. When a particular title has multiple versions available online, Google Scholar lists the several versions and their version-years under a single search result. We then download the earliest version-year as our pre-submission version (ensuring that the earliest version-year pre-dates the final publication year). This process yields 3,406 matched pre-submission to published version pairs. Authors verified the match quality by hand for 275 randomly chosen pairs, ensuring that our matching procedure produced accurate results. Nevertheless, any potential scraping errors should not be correlated with author gender and would therefore add noise to our analyses. 

Both published and early versions of each article were downloaded in the PDF file format, but analyses required parsing these documents into plain text files. To do so we use an open-source machine learning library for restructuring PDF documents, called the GeneRation Of BIbliographic Data, or GROBID. Using GROBID, we read all PDFs into .txt files representing the full text, abstract, introduction, and conclusion/discussion sections of each article.\footnote{Not all articles contain a ``discussion'' section. When they do, conclusion and discussion sections are combined together into a single section we henceforth refer to simply as the conclusion.}

\subsection{Classifying Author Gender \label{subsec: gender}}

Our analyses require that we assign author gender to each article, which necessitates several assumptions. First, because gender identity is not observed, we have to infer each author's gender based on their first name and/or second name, which are gathered from the metadata. We combine 4 major name-to-gender inference services: Gender API, NamSor, Genderize.io, and the Gender R package.\footnote{\cite{santamaria2018} evaluate the accuracy of these services compared to one another and find that Gender API performs best on the authors' test set of 7,076 names with an error rate of 0.0789.} For each of the four gender services, we remove all accented letters from names in our dataset. We then infer a gender for every first and second name in the dataset separately, allowing for authors with both a first and second name to have two potentially different gender predictions. These two inferences are done separately in order to allow authors with an anglicized second name a higher chance of being gendered correctly.\footnote{As \cite{santamaria2018} note, gender inference services perform particularly poorly on East Asian names. A large proportion of our dataset's authors who have a first name that is East Asian in origin also use a second name that is more likely to be correctly predicted by algorithmic gender inference.} Next, to address any inference issues that may arise from small sample size of uncommon names, we throw out any inferences that come from a sample of less than 100. The authors with two gender predictions (i.e those with a second name) are narrowed down to one inferred gender prediction by keeping the prediction with higher probability, adjusting for the name's representation in the gender service's full sample. Finally, the predictions across the four services are narrowed down to one inferred gender per author by keeping the inference with highest probability. By combining four gender inference services and using second names, we hope to achieve the most robust results possible. After obtaining a gender prediction from the name-to-gender services, the authors review the results for accuracy and make manual corrections where errors are noted.\footnote{For instance, the name-to-gender services regularly predicted the first name Andrea to be female. However, Andrea's of Italian origin are often male. Authors manually searched author's websites to make such re-assignments. Specifically if an author's website used a gendered pronoun, we ensured that the assigned gender aligned with that pronoun. Inferences are largely unchanged if we do not make such manual corrections, as they just add noise to our analyses.}
	
Next, we aggregate individual author genders into an article-level gender index. The construct we aim for is the editor's overall perception of the article's gender. One way to do so would be to construct a continuous measure of gender, such as the number of women scaled by the total number of authors. However, the structure of many of our latter analyses require a dichotomous split. Thus, we choose to define articles as female if at least one author is a woman. This choice reflects the rarity in our data of articles with more than one woman on the author team.\footnote{Around 6\% of the observations have more than one female author.} Nevertheless, we recognize that our cut-off is subjective and thus test the sensitivity of our results to (1) using a continuous measure of gender in our primary tests, and (2) performing all analyses on a completely different sample of historical dissertations hosted on Proquest, which are all single-authored. Results are robust to these choices.

\subsection{Descriptive Statistics \label{subsec: descriptives}}
Our sample statistics are presented in Table \ref{table: descriptives}. The first four variables are defined at the matched-pair (pre-submission and published version) article level. The average change in the degree of uncertainty from the pre-submission to published article is 3.35 words/phrases, though there is significant variation in the change in uncertainty, with an interquartile range of 47. Thirty-four percent of the articles are written by at least one female. The authors' weighted average number of citations is seven-thousand, though this number is highly right-skewed. 

Of note, the average time that an article spends in the review process is 2.6 years. The relatively long review process is conventional in the fields in our sample, and it starkly contrasts the much shorter review times noted in the hard sciences.\footnote{For instance, the average peer review duration for articles published in \textit{Science} is 114 days (https://academic-accelerator.com/Review-Speed/Science).} The long review process in our sample is due to multiple ``rounds'' of feedback between authors and editors, whereby editors (and their selected reviewers) typically provide written comments and suggestions for how the authors should revise their paper for the next round. In other words, our setting is prime to detect any potential direct influence that an editor has on the features of a manuscript. 

Table \ref{table: genderdescriptives} provides univariate comparisons of articles' uncertainty indices, split by author gender. In Panel A, we compare the level of uncertainty in the pre-submission versions relative to the published versions. In the pre-submission versions of articles, there is no statistically significant difference in the level of uncertainty between male-authored papers versus female-authored papers. However, a large and statistically significant gender-based difference in uncertainty arises in the published versions of those same manuscripts. Descriptively, the evidence points to a potential role of the publication process itself in contributing to differentially larger increases in uncertainty for female authors relative to male authors.

In Panel B, we split this analysis by our two measures of uncertainty: that coming from the hedge words obtained through the SciBERT model and that coming from our additional 13 words in our ``bank of words.'' We find a consistent story across both of these measures of uncertainty, whereby pre-submission versions are statistically similar among male and female authors, but larger increases in uncertainty arise in women's published manuscripts.

Finally, it is worth noting that the changes in uncertainty may differ based on the placement in an article. Whereas prior literature typically focuses on narrow aspects of an article's text, such as the abstract, we are able to explore uncertainty placement, since we gather and analyze full-texts and split articles into sub-sections. In Panel C, we provide univariate evidence on the changes in uncertainty by section. Interestingly, we find no statistical difference in male and female changes in uncertainty in the abstract, highlighting the importance of more complete analysis of articles as a whole. We find that females have larger increases in uncertainty in both the full text and in the conclusion sections of manuscripts, and that male authors have larger \textit{decreases} in uncertainty in the introduction of the articles.

\subsubsection{Changes in uncertainty by field and journal}
In Table \ref{table: byjournal}, we explore the changes in uncertainty by field and by journal. Prior work highlights the potential heterogeneity in the treatment of women across sub-fields (e.g., \cite{beneito2018}; \cite{chari2017}). Interestingly, we find the smallest (negative) difference in the changes in uncertainty by gender in the field of management; this is perhaps expected due to the higher representation of females in this field, whereby gender bias may play a smaller role in the editing process. The fields of accounting and finance reveal larger increases in uncertainty for females relative to males. Perhaps surprisingly, manuscripts published in the field of economics are statistically similar in the changes in uncertainty for female-written papers relative to male-written papers. 

\section{Methodology \label{sec: method}}
Next, we aim to isolate the role of author gender in differential changes in uncertainty in a multivariate regression analysis that controls for potential confounding variables. Specifically, it's important to recognize that changes in uncertainty may be explained by other characteristics of an article that may correlate with gender, such as the topic, the method, and the quality of execution -- what we henceforth collectively refer to as the articles ``fundamentals.'' Our regression seeks to control for these fundamentals such that the remaining variation in the changes in uncertainty can be causally attributed to the author's gender.

Our regression is estimated by OLS using the following specification:
\begin{equation}
y_{i_{pub-pre}} = \beta Female_{i} + \gamma Controls_{i} + \alpha_{i} + \alpha_{t} + \varepsilon_{i,t}
\label{eq: main}
\end{equation}
where $y$ is the change in the uncertainty words and phrases (captured by both hedges and the additional bank of words) for an article, \textit{i} from the pre-submission version to the published version (\textit{pub-pre}). In other words, each observation represents a matched pre-submission to published version pair, and the dependent variable captures the change in that pair over time. \textit{Female} is an indicator variable equal to one if the article is authored by at least one female. We include several control variables to capture variation in topic, quality of execution, and other fundamental characteristics. \textit{Citation Index} is an equal-weighted average of each author's citations (in thousands), obtained from Google Scholar as of March 2022.\footnote{We would ideally measure citations as of the point of article publication, but retrospective data is not available. Citations were gathered by building a scraper for each author's GS page, which lists the citation index.} We also control for the time, in years, between the pre-submission version and the published version, since articles spent longer in review are expected to undergo larger revisions. Because we cannot perfectly observe the date and the version of the first submission, our variable, \textit{Review Time} captures the time that an article spends in the review process with noise. In all specifications, we include journal and (published) year fixed effects.

It is important that we control for research topic, as prior work shows that men and women choose different specialties (\cite{chari2017}). We use Latent Dirichlet Allocation (LDA), an unsupervised clustering method to categorize documents into a selected number (N) of topics. LDA assumes that (1) Documents fall into topics in a Dirichlet distribution; and (2) N topics fall into word space in a Dirichlet distribution. These assumptions mean a single document is made up of 1 to N topics, and each topic is represented by some collection of words. Given that our articles come from 15 different economics journals and form a sample of 3,406, we believe this is a reasonable assumption. At a high-level, LDA takes advantage of the idea that documents that have very strong commonality in their vocabulary patterns are likely to have the same dominant topic(s) and/or methods. Using our data, the full text of each published article is cleaned, tokenized, and lemmatized into a list words greater than one letter using Python NLTK. We also add any bi and trigrams that appear twenty or more times in the corpus. The final bag of words is made up of words, bigrams, and trigrams that appear in at least twenty articles but no more than 50\% of articles, in order to exclude rare words and stop words, respectively. We use the Python library, Gensim, to create an LDA model of 25 topics, and then extract the dominant topic of each article to include as a fixed effect in all regressions. The output of our LDA topic modeling can be viewed in Table A1 of the Appendix.

\section{Baseline Results \label{sec: results}}

Results of our primary regression are reported in Table \ref{table: primary}. In columns 1 through 4, we estimate separate regressions for each section of the manuscripts. We find that manuscripts written by women increase their uncertainty by 5.3 more words/phrases than those written by men, when we consider the full text of the documents. This result holds after controlling for citations, review time, and journal, topic, and year fixed effects. Relative to the baseline number of uncertain words in pre-submission versions of manuscripts, the estimate indicates about a 3\% larger increase in uncertainty for women versus men.

In the multivariate regression results, we do not find statistically different changes in uncertainty between men and women in abstracts nor in introductions. However, column 4 reveals a statistically larger increase in uncertainty for females in the conclusion section of the manuscript, although the magnitude is modest. In column 5, we re-estimate the regression for the sample of full text articles, while adding a control variable for the change in the total number of words from the pre-submission version to the published manuscript. On the one hand, the change in the total number of words may represent changes in fundamental aspects of the paper (in which case we would want to control for this); on the other hand, changes in the number of words may represent gender-specific biases (e.g., requiring females to add more analyses relative to men). Thus, we believe that including the change in total words as a control variable yields very conservative estimates of the effect of gender on uncertainty changes. Results reported in column 5 remain significant, though the magnitude is somewhat muted. 

One may be concerned that our measure of quality -- citations -- only imperfectly measures our construct of interest. For instance, since on average men hold more senior positions they may be more ``skilled'' than women in our sample, and this level of experience may not be fully captured by citations. We further control for skill based on the author's publication number -- i.e., whether it is their first publication, second, etc. Specifically, for each author in the sample, we sort their publications based on date published and assign an ordering (1st, 2nd, ... 5th) to each publication. Because of sample attrition concerns, we require each author to have at least four publications in the sample. Figure \ref{fig:pubnumber} presents visual evidence of changes in uncertainty across publication numbers. One can see that within each level of experience (publication number), females exhibit larger increases in uncertainty relative to males; further, the difference gets \textit{larger} over time/experience rather than smaller. Evidence is inconsistent with an omitted ``quality'' factor and suggests that effects do not dissipate as women become more experienced.

Finally, we note that our measure of uncertainty is slightly skewed, thus a log transformation is desirable. Because the change in uncertainty can take on negative values, to preserve these values we take the log-modulus transformation, or the signed log of the absolute value of \textit{$\Delta$ Uncertainty}.\footnote{In sensitivity tests, we also use a cube-root transformation, and results are similar.} Results, reported in column 6, are consistent with other specifications, and suggest that papers written by female authors exhibit about a 12\% larger increase in uncertainty relative to papers written by males.

\subsection{An alternative approach: NLP Modeling for causal analysis}
The key quantity we want is the natural direct effect (NDE) (\cite{pearl2001direct}) of gender on the usage of uncertainty expressions. However, there are many potential mediators that we need to control for, and though our baseline analyses presented in Section \ref{sec: results} are comprehensive, one may still be concerned about confounders. To this end, we train an NLP model to simulate the counterfactual, ``\textit{had all of the fundamental variables been the same, how would the usage of uncertainty expressions differ for male and female authors.}'' We build an NLP model to take as input all the fundamentals, including the text with deleted uncertainty words, and non-textual meta information such as author genders, citations, number of authors, year of the publication, number of sentences and words of the paper, and sentence length and placement.

The inner working of our NLP model is illustrated in Figure \ref{fig:concat_bert_arch}. Specifically, to model the text, we adopt the general-purpose SciBERT model to encode the text into a 768-dimensional vector to capture the fundamental features encoded in the text, such as the semantics, writing style, and vocabulary. To avoid overfitting, we add a dropout layer on top of this high-dimensional text vector (\cite{srivastava2014dropout}). To model the non-textual variables including gender, we encode them as a vector, and concatenate it with the text vector. With the combined vector which encodes both gender and all the fundamental information of the writing, we use two layers of fully-connected neural networks to process the features, and finally fit them for the regression target, namely to predict the number of uncertainty expressions.\footnote{We open-source our codes at {\url{https://github.com/causalnlp/ai-scholar-uncertainty-prediction}}.}

\subsubsection{NLP Modeling: Implementation Details}
We tokenize each full text article into sentences using the NLTK Python package (\cite{BirdKleinLoper09}). Every sentence has an uncertainty score calculated by adding the number of hedges and dictionary words within that sentence. The remainder of the NLP modeling sections will refer to an observation as a single published sentence rather than an article. Sentence tokenization yields a total of 1,897,963 sentences used in our NLP modeling. Since our goal is an uncertainty language prediction task using just text fundamentals, we keep the main context by removing all detected hedges and uncertainty-related words. Removal of uncertainty language ensures that our model does not act as a simple counter of the hedge and dictionary words.

In addition to the features used in earlier article-level analyses, we create several new variables derived from the relationship that sentences have with each other and their respective articles. For every sentence, we calculate the placement of the sentence in its corresponding article (1st, 2nd, ... $n$-th) as well as the number of words in the sentence. Every article also has an average sentence length and number of sentences. As such, sentences from the same article have the same value for average sentence length and number sentences.

For the training details of our model, we split the dataset into training (using 80\% of the entire data), development (10\%) and test sets (10\%). We use the training set to finetune the NLP model, use the development set to select the best hyperparameters, and report the final performance on the test set. When searching for the best hyperparameters for our SciBERT-based model, we conduct a grid search over batch sizes of 8 and 16, learning rates of 5e-6, 1e-5, 5e-5, and 1e-4, dropout rates of 0.1 and 0.3, as well as the dimensions of the fully-connected layers of 50 and 100. We identify the best model with a batch size of {8, learning rate of 1e-5, dropout rate of 0.1, and fully-connected layer dimension of 50}. 
We initialize the pretrained SciBERT model with the \texttt{scibert\_scivocab\_uncased} checkpoint from the transformers Python library (\cite{wolf2019transformers}), and we finetune the entire model for {three} epochs to reach the convergence.

\subsubsection{NLP Modeling: Results}
Our model achieves 0.8080 R2 scores on the test set. Since the number of uncertainty expressions in sentences follow a long-tail distribution, we visualize the model results for 0 to 3 uncertainty expressions. Figure \ref{fig:heatmap} shows a heatmap of the confusion matrix between the true number of uncertainty words and the predicted number. Most of the numbers concentrate on the diagonal, showing that, for majority of the times, our model correctly predicts the number of uncertainty words given the text and meta information.

Using our powerful NLP model that has learned the role of the study's fundamentals and the author's gender in determining uncertainty, we can now answer the counterfactual question, ``\textit{holding fundamentals constant, how much more uncertain would a counterfactual female-written article look, relative to the male-written article?}''. We extract all the articles written by male authors, feed them into our NLP model, and artificially `flip' the gender from male to female. This allows us to construct an all-else-equal ``pseudo-female'' observation. Comparing the real male change in uncertainty to the pseudo female change in uncertainty produces a distinct difference that is consistent with our primary finding: Table \ref{table: pseudo} shows a mean value of 201.79 uncertainty expressions per male article, relative to 207.53 (+5.74) uncertainty expressions per counterfactual female article.

\section{The Role of the Editor \label{sec: editor}}
Having established that women-authored papers exhibit larger increases in uncertainty relative to male-authored papers, we next move to understanding whether the editorial process contributes to this pattern. Specifically, we hypothesize that the editor in charge of the manuscript during the review process will have a measurable effect on the paper's outcome, including how the inferences of the paper are framed. There are several reasons and possible mechanisms through which an editor influences the voice of a paper. First, as noted in Section \ref{subsec: descriptives}, papers spend a significant length of time in the review process and go through multiple rounds of revisions; during that time editors and reviewers make suggested edits and authors respond to those suggestions. Thus, there is ample time and input from the editorial team to influence a paper. Second, in each round editors typically provide written comments to the author (see Table A2 in the Appendix for an example), where they can make strong suggestions for how the paper should be framed. Third, editors may play an indirect role in how a paper is framed through their selected review team (i.e., associate editors and reviewers); selecting ``tougher'' reviewers for women-authored papers may result in the pattern of evidence we document.

\subsection{Identifying the Editor}
Authors often acknowledge the editor in charge of their manuscript in the footnotes of the text.\footnote{Most often the footnote on the title page, though there is significant variation across manuscripts.} Editor acknowledgement is voluntary on the part of the author, however it is conventional practice for many journals. Extracting the name of a paper's editor was a non-trivial task, as articles are not consistent in whether they name an editor, how they name the editor, and where in the paper the editor is named. In order to ensure all named editors were collected, we complete the editor extraction process in a combination of automated and manual steps. First, we use Python to extract portions of text where an editor may be named, first parsing through footnotes and then through the full text. Within each extracted text pattern, we implement the Spacy Python Package, using the named entity recognition software to select a potential editor name. The automated process produced editor names for 2,357 articles. Our automated extraction was imperfect; thus, to avoid missing observations we manually read footnotes for the remaining articles in order to extract any missing editor names. This manual process added another 141 articles with editor names. 

\subsection{Estimating an ``editor effect''} 
Our objective is to identify the role of specific editors in contributing to the larger increase in uncertain language for female authors, relative to male authors. In other words, after controlling for the fundamental characteristics of a manuscript, does the impact of the author's gender on his or her changes in uncertainty depend on who they are assigned to as editor of the paper? The empirical tactic we take is to regress the change in hedge words on a vector of control variables, the author's gender, editor fixed effects, and the interaction between editor fixed effects and the author's gender. We can then assess the importance of the editor in explaining the author-gender gap by examining the explanatory power and joint significance of the editor*female fixed effects.

Our data are particularly well suited to isolate and extract an `editor effect' due to the network of author-editor relationships, whereby authors have multiple papers under review (often by different editors), and editors oversee papers written by a host of different authors. This network allows us to utilize a rigorous specification that simultaneously controls for author fixed effects and editor fixed effects. To do so, we expand the dataset so that it is at the author-article level. In other words, if Article A has three authors, we expand the single article-level observation to three article-author level observations: one for each author. Doing so allows us to include an author fixed effect which importantly controls for the author's time-invariant specific writing style, approach, and quality. Similarly, we hold the editor's style (e.g., relative strictness) constant through the editor fixed effect.  Under this fixed effect structure, we can then measure how each editor influences female-written papers relative to male-written papers. We specifically estimate the following equation using OLS:
\begin{equation}
y_{i,j,a,t_{pub-pre}} = \alpha + \beta Controls_{i,t} +  \sum \delta_{j} E_{j} + \sum \gamma_{j}  E_{j} * Female_{i,j,a,t} + \alpha_{a} + \alpha_{t} + \varepsilon_{i,j,a,t},
\label{eq: second}
\end{equation}
where the dependent variable is the article's change in uncertainty cues, $E_{j}$ are editor specific fixed effects, \textit{i} is an article index, \textit{j} is an editor index, \textit{a} is an author index, and \textit{t} is a time index. $Female_{i,j,a,t}$ is defined at the individual author level (rather than at the author-team level in the article-level analyses), control variables include the time the article spent in review, the changes in the article's total words, and fixed effects include author, editor, year published, topic, and journal, in the most rigorous specifications.

Table \ref{table: editorfe}, Panel A reports our results. The specification reported in the first row excludes editor fixed effects and the associated interaction term with the author's gender. Specifically, the specification regresses the change in uncertainty cues on author fixed effects, the time the article spent in review, the total change in words, and journal, topic, and year fixed effects. The Adjusted \(R^{2}\) for this specification is 68 percent. In the second row, we add editor fixed effects and their associated interaction with the author's gender. The adjusted \(R^{2}\) for this specification increases substantially, to 71 percent. This suggests that the identity of the editor is important in explaining the author-gender gap. Further, the F-test for the joint significance of the editor*female fixed effects is large, leading us to reject the null hypothesis of no joint effect.

Though we can control for time-invariant author and editor characteristics, it is worth noting that author-editor matches are not always random. In fact, many articles are assigned to editors based on a match between the topic and the expertise of the editor. Thus, we must be cautious in the interpretation of the results in Panel A of Table \ref{table: editorfe}, as endogenous matching may confound our inferences. We do point out, however, that any omitted `matching' variables would have to systematically line up with author gender, (e.g., editor's exhibiting larger author-gender gaps are systematically matched with females' lower quality topics but males' higher quality topics), and not be subsumed by our control variables (e.g., LDA topic).

To ease concerns about endogenous matching, we would ideally randomly assign papers to editors. We sent an email to the managing editor of each journal to inquire about how they assign papers to editors. Emails can be viewed in Appendix A3. Two journals, the Journal of Accounting and Economics and the Journal of Accounting Research, responded that assignment was random. For these two journals that randomly assign manuscripts to editors, we test for the joint significance of editor*female fixed effects and report the results in rows 3 and 4 of Table \ref{table: editorfe}, Panel A. For the subset of randomized editors, we continue to reject the null of no joint effect, suggesting that omitted matching variables do not fully explain our results.

Table \ref{table: editorfe}, Panel A shows that editor*female fixed effects explain a substantive portion of the variation in changes in uncertainty, and that collectively, the fixed effects are statistically different from zero. Next we assess how different the editors are in their author-gender gap (i.e., the inter-editor author-gender gap). Thus, we report the distribution of editor*female fixed effects, estimated from equation \ref{eq: second}. Table \ref{table: editorfe}, Panel B reports the standard deviation of the fixed effects, and the 25th, 50th, and 75th percentiles. In computing these statistics, we extract each editor's editor*female fixed effect, $\gamma_{j}$, from estimating equation \ref{eq: second}, and weight each fixed effect by the inverse of the square of the standard error to account for estimation error. The distribution indicates that the magnitude of the inter-editor author-gender gap is economically large; the difference between females and males assigned to an editor in the 75th percentile versus females and males assigned to an editor in the 25th percentile is about a 53 word, or 26\%, larger female penalty.\footnote{Percentage magnitudes are calculated relative to the benchmark pre-submission average number of 204 uncertainty words.}

A concrete example best illustrates these results. Assume that authors Sally (female) and Jack (male) are assigned to an editor in the 25th percentile of the author-gender gap distribution, while authors Jill (female) and Steven (male) are assigned to an editor in the 75th percentile of the author-gender gap distribution. Aside from the author's genders, assume that their manuscripts are equivalent on all fundamentals. The distribution shown in Table \ref{table: editorfe}, Panel B suggests that the difference between Jill and Steven's change in linguistic uncertainty is 53 words larger than the difference between Sally and Jack's change in linguistic uncertainty, simply based on who they were assigned as their editor.

\subsubsection{Correlating the editor effect with editor characteristics}

Due to the substantial heterogeneity in editor*female fixed effects, we next explore whether any observable traits of the editor correlate with their author-gender gap. To do so, we construct a dataset of editor*female fixed effects ($\gamma_{j}$) and regress these fixed effects on editor-level characteristics. Since there are 88 editors in the sample, our regressions include 88 observations, where each observation is weighted by the inverse of the square of the standard error.

The traits we explore capture the editor's demographic characteristics as well as their personal experiences, as prior literature has found that these characteristics often influence how individuals perceive and act on gender biases (e.g., \cite{guiso2008}; \cite{abrams2012}; \cite{siniscalchi2021}; \cite{benson2021}). First, we follow \cite{guiso2008} and investigate variation in gender equality indices from the editor's country of origin. Specifically, cultural inequalities vary significantly across countries, and those ingrained cultural norms may influence how an editor perceives women versus men. We obtain an editor's country of origin either from their CV (which often states citizenship an/or birth country), their website, or Wikipedia. In cases where citizenship differs from birth country, we take birth country as the editor's country of origin. As our measure of gender equality, use the 2021 Global Gender Gap Report, which constructs a Gender Gap Index (GGI) for each country based on economic and political opportunities, education, and well-being for women. The GGI is produced on a scale of 0 to 100, where a higher score indicates more gender parity.

The second editor characteristic we explore is their gender, as \cite{siniscalchi2021} argue that self-image bias may influence how evaluators assess their subordinates. One such salient trait is the editor's gender, whereby male editors may favor male authors and female editors may favor female authors. Editor gender is inferred from both name and personal pronouns used on the editor's website.

Third, we explore the editor's prior work experience -- specifically, who he or she works with as coauthors in their own studies -- as an indication of bias. There are many reasons to believe that working with a larger proportion of men may influence (conscious or unconscious) views on women. First, \cite{harris2019} show that prior work experience indicates a judges ideology and influence his or her sentencing patterns. Further, surrounding ourselves with similar types of people may exacerbate confirmation biases and therefore views toward minority groups (women, in this case). We explore an editor's work experience by scraping their Google Scholar page for each published article and the associated list of co-authors. We then use name-to-gender services to code each co-author's gender, and we construct an index of male co-authorships. Specifically, for each paper we calculate the proportion of men to total coauthors; we omit the editor himself in this count so as not to disproportionally penalize editors who are men. We then take an equal-weighted average of the index across all of the editor's published articles.\footnote{This task was performed by hand (from the editor's CV) for those editors who do not have a Google Scholar page.}

Finally, we explore the role of the editor's PhD graduation year in shaping their author-gender gap. There is some evidence that gender biases have declined over time. Further, \cite{schoar2016} shows that their MBA graduation year influences how executives manage and perform within their organizations in the future. We gather the editor's PhD graduation year from each of their CVs.

Descriptive statistics on each of these variables are reported in Table \ref{table: descriptives}. The average Gender Gap Index (in percentage terms) is 74, suggesting that the average editor in our sample is from a country with a relatively high GGI score. As expected, most of the editors in the sample are men (84\%), and the editors in the sample have co-authorships that are disproportionally represented by male authors (84\%). Finally, the mean PhD graduation year is 1994, though there is significant variation.

Graphical evidence is reported in Figure \ref{fig:editorch}, and we report the correlation matrix and the results of a regression of the editors author-gender gap on his or her characteristics in Tables \ref{table: corrmain} and \ref{table: editorFEmain}, respectively. Three of these characteristics have a statistically and economically significant relationship with the editor*female fixed effects. In particular, editors from countries with a higher GGI index (indicating more gender parity) exhibit a lower author-gender gap; a one percent increase in GGI reduces the gap by 2.5 words. Editors who have a history of working with a higher proportion of male authors have a higher author-gender gap, and editors who graduated with their PhD more recently have a lower gender gap. Interestingly, the relationship between editor gender and the author-gender gap is statistically insignificant; however, we should note that there are very few female editors in our sample.

While the results in Table \ref{table: editorFEmain} are not causal, they provide suggestive evidence that an editor's personal traits influence how he or she treats women authors versus male authors in the review process.

\section{Robustness \label{sec: robust}}
Our results thus far show that the author-gender gap lines up with the identity of the editor. In other words, it matters who you get assigned as your editor. However, interpretation is nuanced. For instance, it may be the case that editors actively treat women authors different than men authors (e.g., through different guidance provided in editor letters). It may also be the case that certain editors are more critical -- to both female and male authors -- than other editors, but that female authors \textit{are more sensitive to criticism than male authors}. In other words, women are more reactive to negative editors than men are. To provide some preliminary evidence to this end, we sort editors into ``harsh'' editors versus ``lenient'' editors. Harsh editors should be more strict on \textit{both} male and female authors, rather than just female authors. They may also have a lower article acceptance rate, relative to more lenient editors. 

We construct a measure of an editor's harshness on male authors -- in particular, the editor*male fixed effect.\footnote{Note that this is the main effect of the editor fixed effect, extracted from equation \ref{eq: second}.} Generally harsher editors should show larger increases in uncertainty for male authors and female authors alike. Table \ref{table: robust} reports the correlation between the editor's effect on female authors versus the editor's effect on male authors; the two measures are not statistically significantly correlated. 

As a second measure of editor strictness, we construct a proxy for their paper acceptance rate. While we cannot directly observe the portion of papers that they accept or reject, we measure this noisily by taking the total number of papers they accepted during the sample period, relative to the total number of papers that the journal published. Under the assumption that stricter editors have lower acceptance rates, a statistically significant negative correlation between acceptance rates and the editor*female fixed effect may indicate that female authors have stronger reactions to harsh editors. The relationship reported in Table \ref{table: robust} is statistically insignificant.

Finally, it is worth noting that our sample is highly selected, in that we require each manuscript to be published, and we look at how those published manuscripts changed over time. A perhaps more critical decision is whether a manuscript gets rejected from the journal. Should a gender-biased editor also use discretion at the accept-reject margin, we believe the overall impact of an editor on female-authored outcomes is understated in our analyses.

An alternative story may be that editors lower the bar for female-authored papers in the reject-accept decision, which would impact the relative quality distribution in our sample. Accordingly, if females in the sample have lower quality papers than males, then the relatively larger increase in uncertainty may be warranted by fundamentals. We note that this concern would arise if our analyses do not perfectly control for variation in fundamentals, which we have gone to great lengths to do. Nevertheless, we further test the validity of this concern by calculating a measure of the editor's female-specific acceptance rate. Editors with higher female acceptance rates -- should those higher rates be driven by a lower bar for female authors -- should then exhibit higher author-gender gaps. Evidence in Table \ref{table: robust} does not support this alternative story, alleviating such concerns about sample selection.

\section{Discussion and Conclusion \label{sec: concl}}
We document robust differences in the changes in linguistic styles for female authors relative to male authors over the course of the academic review process. The differential gender-based changes persist after controlling for a host of fundamental characteristics of the study and after employing an NLP model that creates all-else-equal male-authored and female-authored counterfactual manuscripts. Our analyses point to an important role played by the editor in gender-based linguistic disparities. We find considerable heterogeneity across editors in how they impact female authors versus male authors -- i.e., their author-gender gap. Further, we show that the editor in charge of the manuscript plays an important role in linguistic disparities even when manuscripts are randomly allocated to editors.

Our analysis further shows that variation in editors' gender-gap indices is systematically related to editor characteristics, whereby editors that are from countries with lower gender parity, those who work with more men as coauthors, and those who graduated in earlier years exhibit higher gender-gap linguistic disparities than editors who are from countries with greater parity, who work with more women, and who graduated with their PhD more recently. Although these additional results are not causal, they begin to shed light on the influential role of editor characteristics in the academic review process.

Editors are critical gatekeepers to the academy, whose decisions ultimately shape the composition of our academic institutions. Given that there is rarely one objectively correct editorial decision, it is not surprising that decision patterns vary across editors. Yet, we are unaware of any prior study that investigates the direct role of individual editors in shaping authors' outcomes. Our study is a first step to shed light on this important role and potential biases that may arise.

As an important caveat, we cannot speak directly to the mechanism through which editors influence authors' outcomes. For instance, the author-gender gap may arise through either overt or implicit biases, which guide how an editor shapes the comments and suggestions brought forth in the editorial process. Alternatively, the editor may play an indirect role -- for instance by choosing different types of associate editors and reviewers for female authors versus male authors. Finally, editors may have specific styles (e.g., `critical' versus `encouraging') that differentially impact female authors relative to male authors, although evidence in Table \ref{table: robust} does not generally support this hypothesis. It is beyond the scope of our paper to sort out these various mechanisms, and this is a task we leave to future research. In particular, we advocate for greater oversight and transparency into the editorial process, perhaps with a more detailed investigation of editorial letters and reviews.

\newpage

\bibliographystyle{ecta}
\bibliography{cfjm}

\setcounter{figure}{0}
\newpage
\section{Figures}

\begin{figure}[ht]

    \caption{\textbf{Overview of the SciBERT model}}
        \centering
    \includegraphics[width=\textwidth]{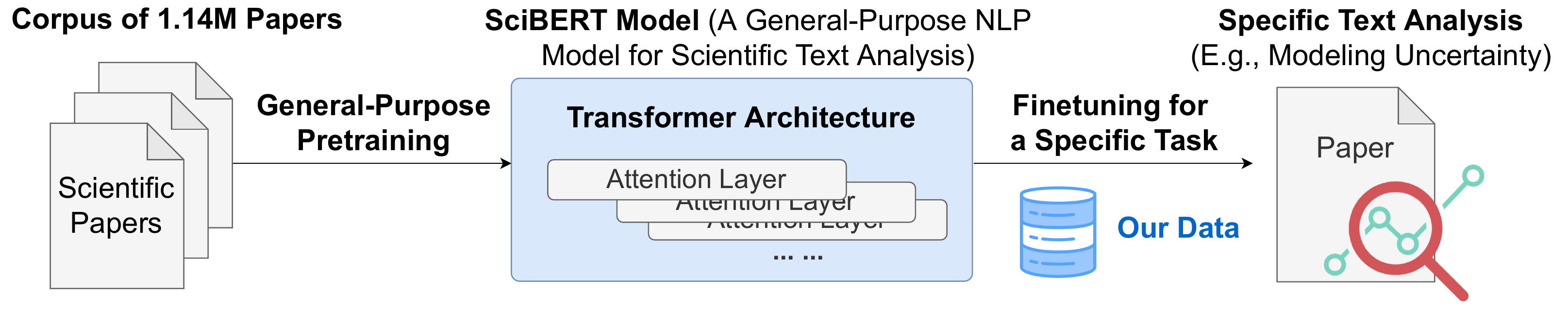}
    \label{fig:intro_scibert}
\end{figure}

\newpage
\begin{figure}[ht]
\caption{\textbf{Change in Uncertainty by Publication Number}\\
This figure presents the evolution of uncertainty changes, split by the ordering of the author's publication number. Results are presented for male authors and, separately, female authors. On the x-axis, we plot the author's publication number, and on the y-axis, we plot the number of uncertain words. Arrows indicate the direction of the uncertainty change, from the pre-submission version to the published version of the manuscript.} 
\label{fig:pubnumber}
\centering
\includegraphics[scale=0.8]{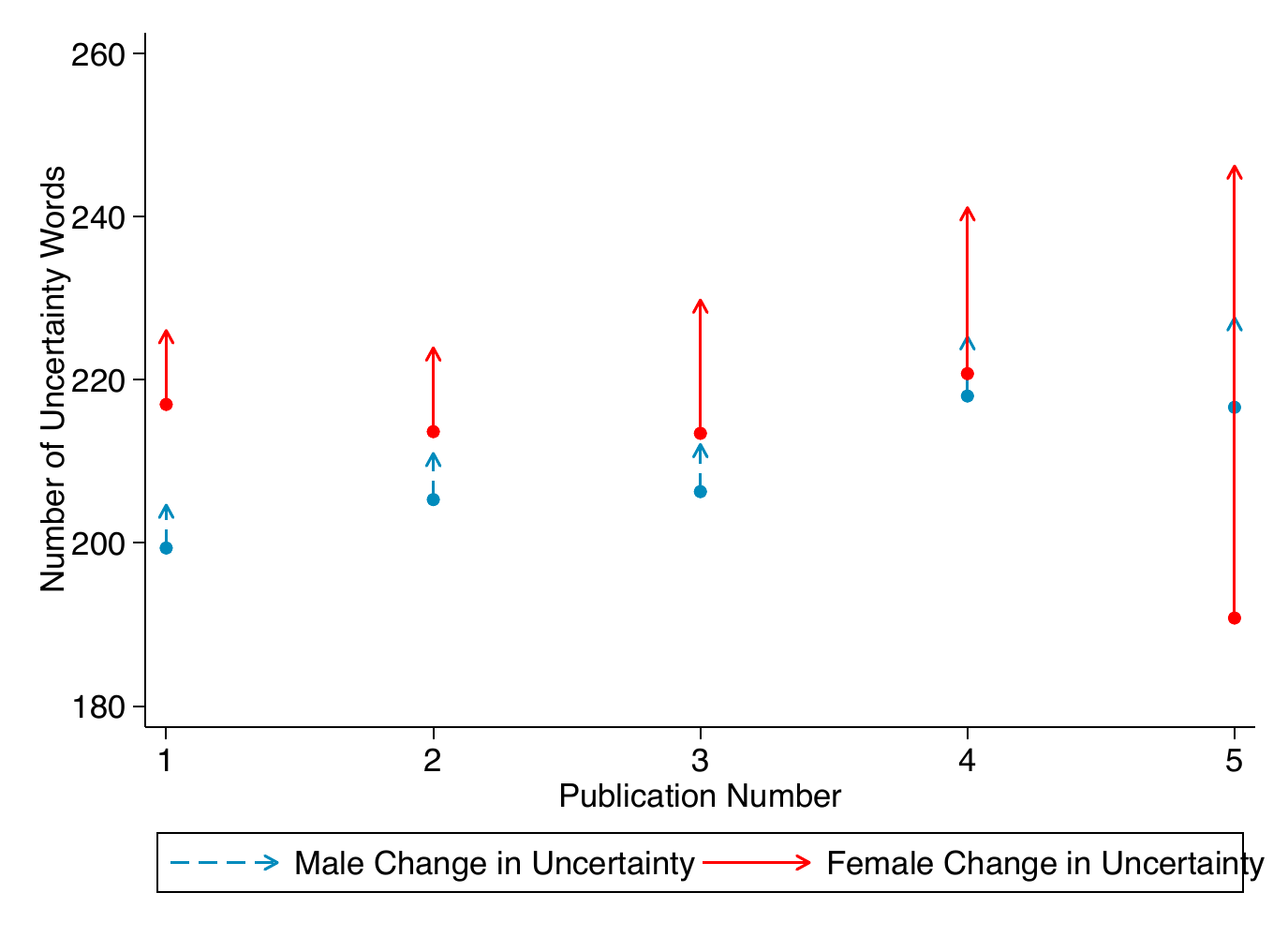}
\end{figure}

\newpage
\begin{figure}[ht]

  \caption{\textbf{SciBERT-based model to predict the number of uncertainty expressions from text and meta features.}\\
  This figure presents a visual representation of the model architecture used in our SciBERT model to predict the number of uncertainty expressions. Inputs include both text and non-text meta features.}
    \centering
  \includegraphics[width=\textwidth]{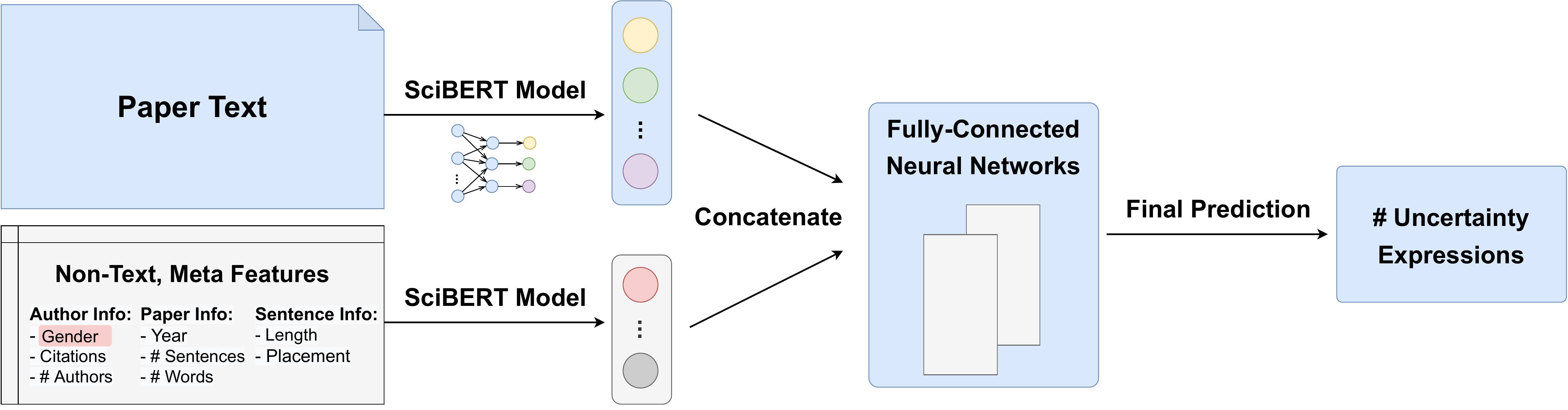}
  \label{fig:concat_bert_arch}
\end{figure}

\newpage
\begin{figure}[ht]
  
  \caption{\textbf{Heatmap of the confusion matrix between the true number of uncertainty expressions and the predicted number.}\\
  Below is a heatmap showing the relation between the true number of uncertainty words and the predicted number of uncertainty words. Due to the long-tail distribution, we visualize the cases with 0 to 3 uncertainty expressions. The large numbers on the diagonal shows that our SciBERT-based model effectively learns the regression task.}
  \centering
  \includegraphics[width=0.6\columnwidth]{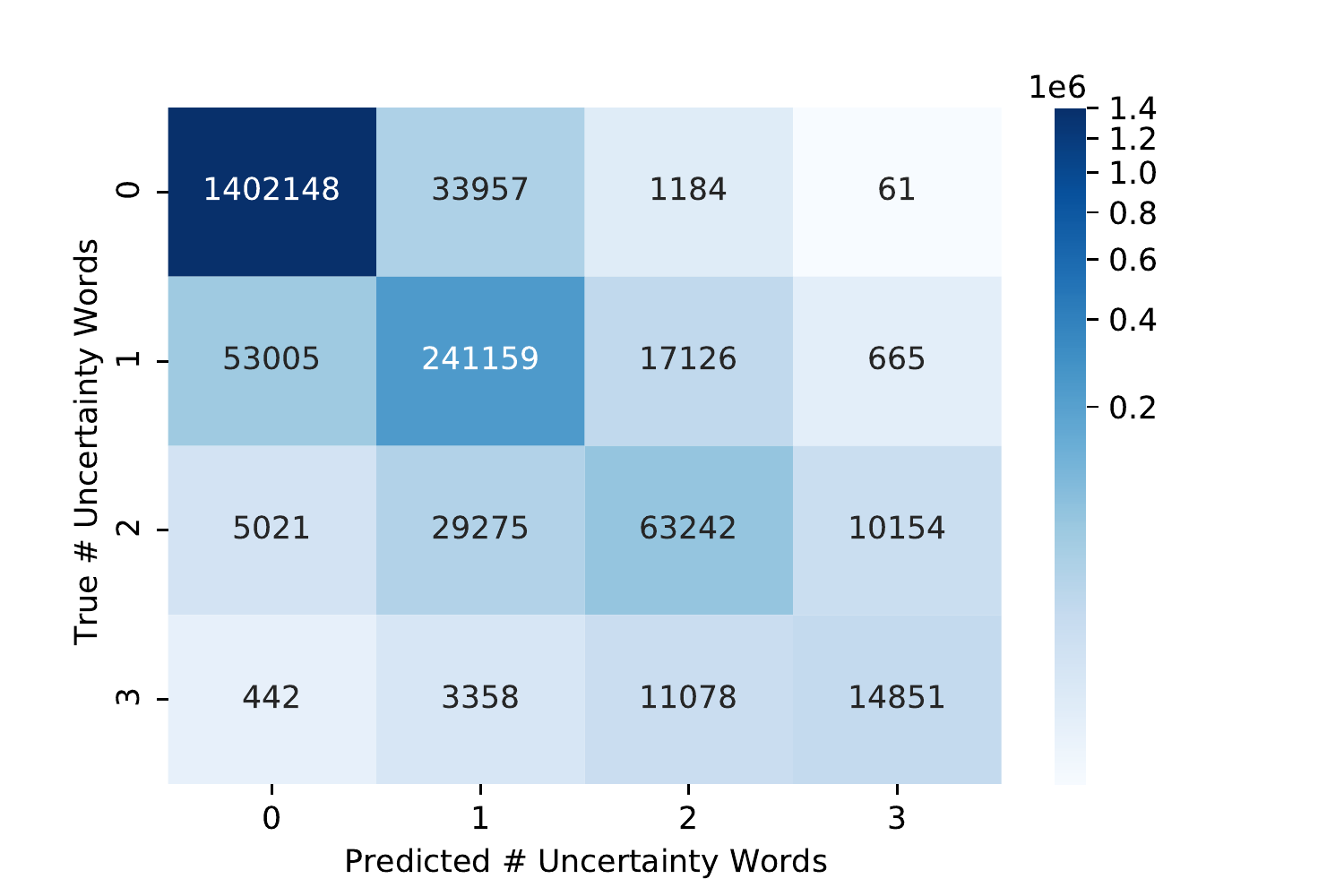} 
  \label{fig:heatmap}
\end{figure}

\newpage
\begin{sidewaysfigure}
\caption{\textbf{Author-gender gap as a function of editor characteristics\label{fig:editorch}}\\
This figure displays correlations between the editor's author-gender gap (e.g., the editor*female author fixed effect extracted from equation \ref{eq: second}) and four different characteristics of the editor. Figure A explores the editors' author-gender gap as a function of the GGI in their country of origin; Figure B shows the correlation with editor gender; Figure C shows the correlation with editors' percentage of male coauthors; and Figure D shows the correlation with editors' PhD graduation year.}
\centering
\begin{subfigure}{0.4\textwidth}
 \caption{Author-Gender Gap by Editor's Country GGI}
    \includegraphics[width=\textwidth]{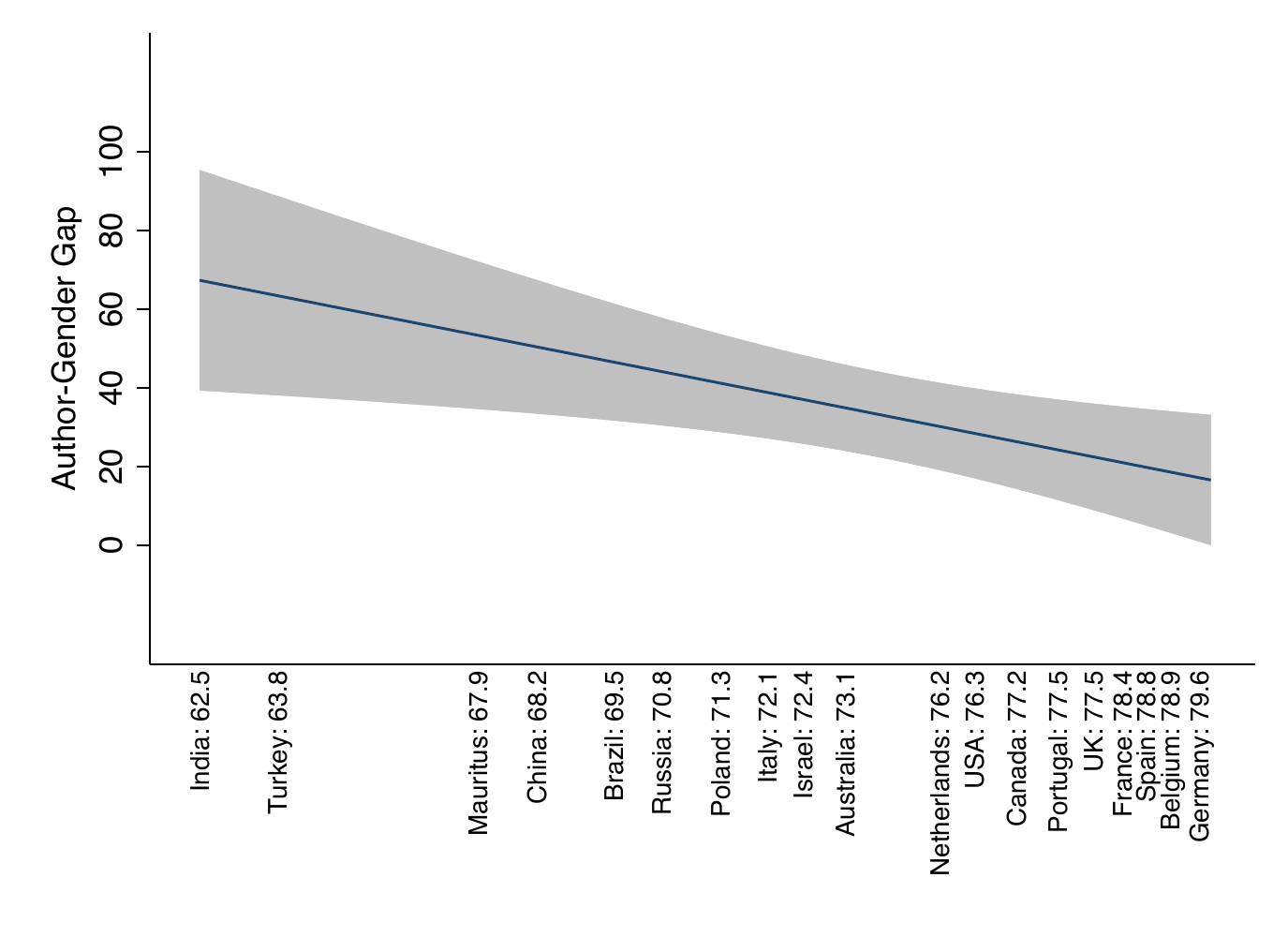}
\end{subfigure}
\hfill
\begin{subfigure}{0.4\textwidth}
 \caption{Author-Gender Gap by Editor's Gender}
    \includegraphics[width=\textwidth]{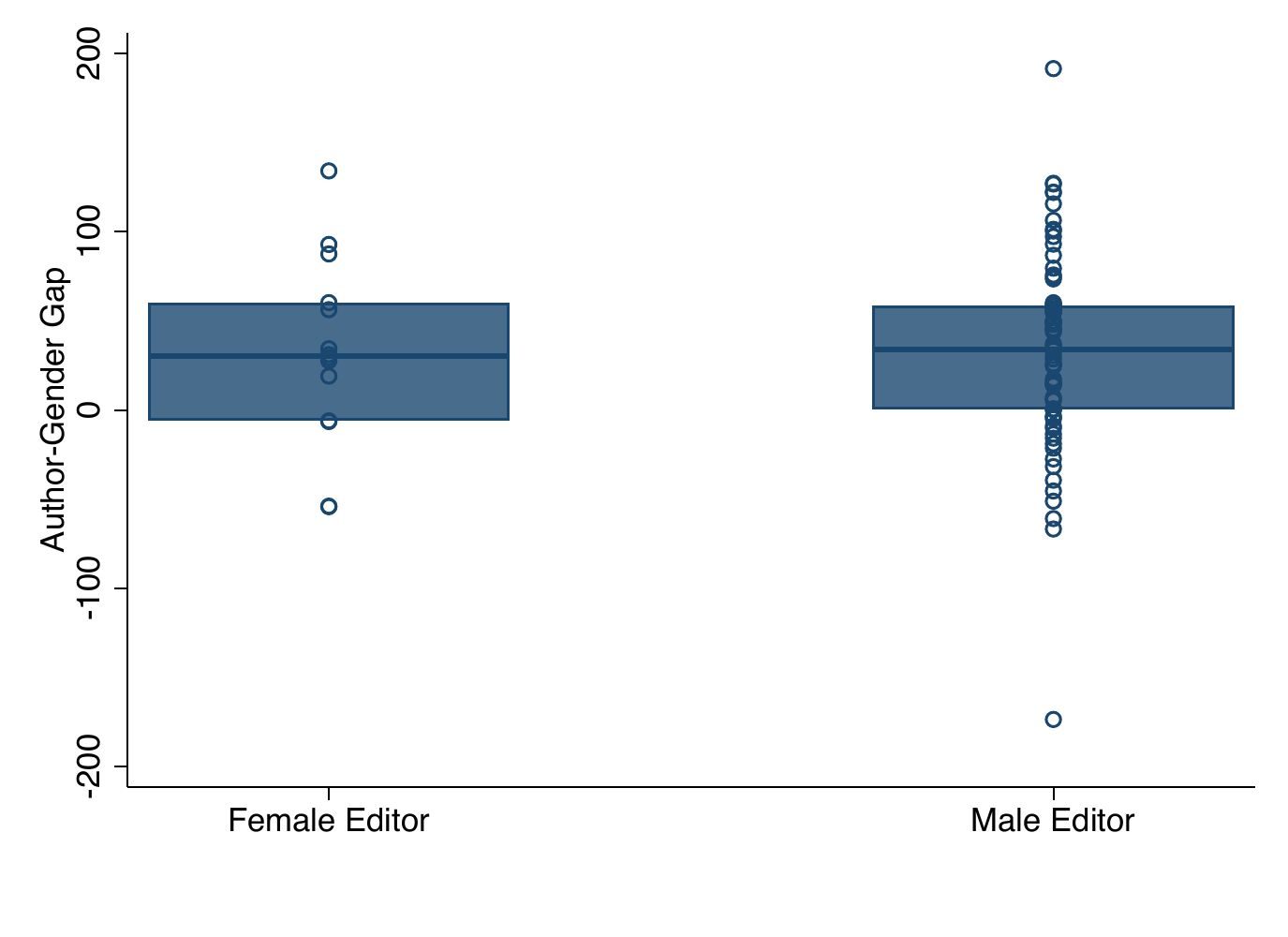}
\end{subfigure}
\hfill
\begin{subfigure}{0.4\textwidth}
 \caption{Author-Gender Gap by Editor's \% Male Co-Authorships}
    \includegraphics[width=\textwidth]{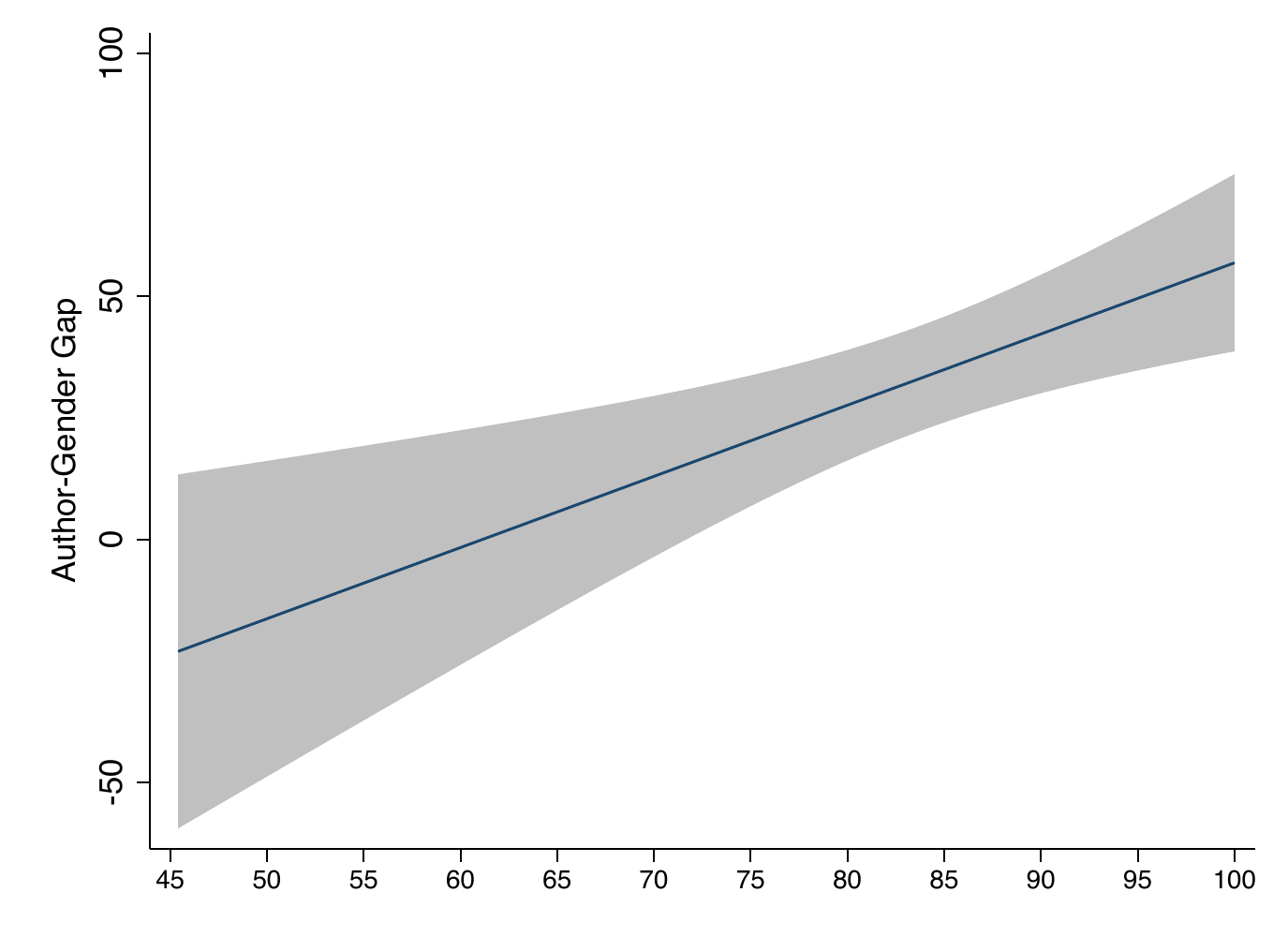}
\end{subfigure}
\hfill
\begin{subfigure}{0.4\textwidth}
 \caption{Author-Gender Gap by Editor's PhD Grad Year}
    \includegraphics[width=\textwidth]{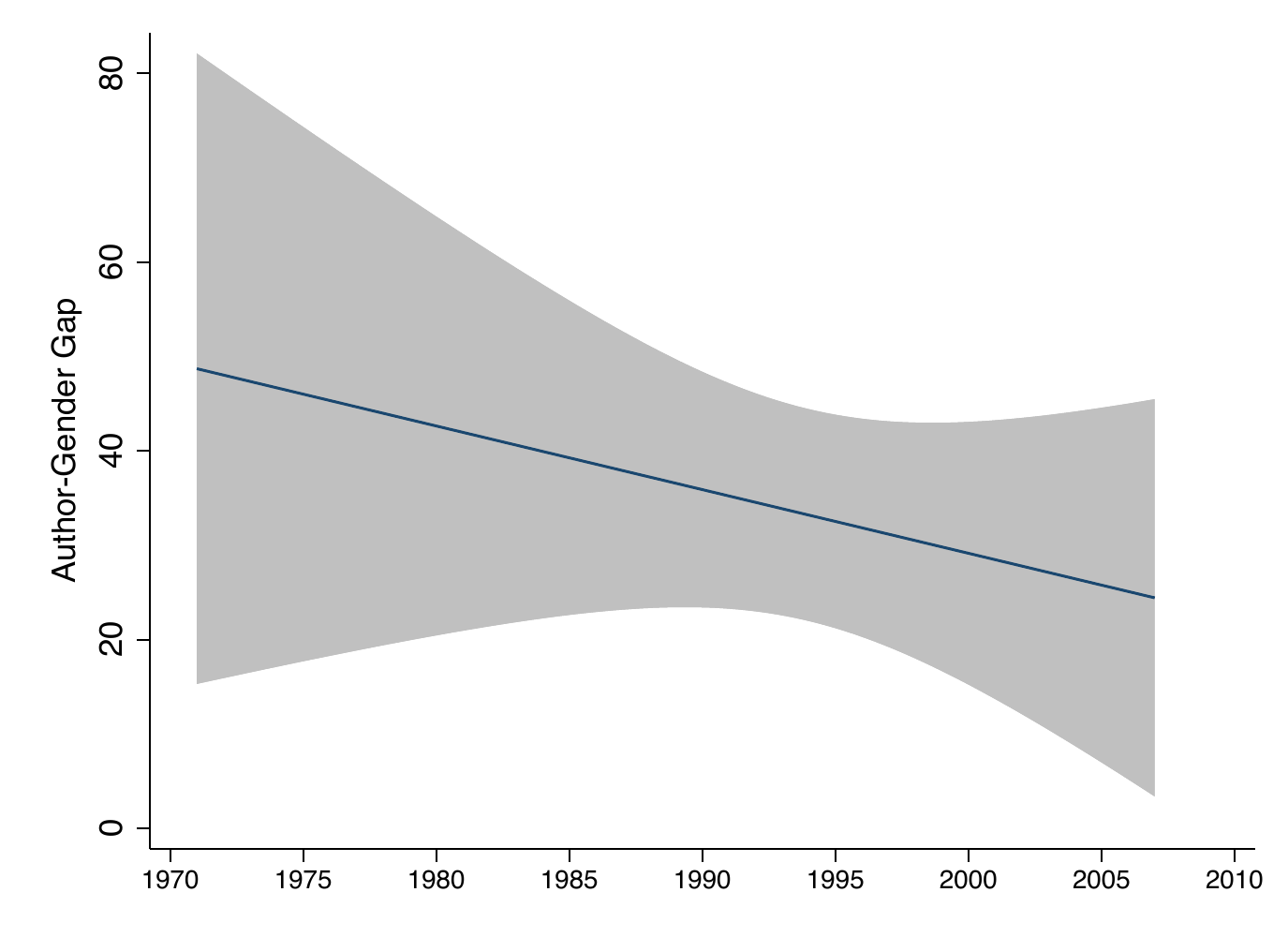}
\end{subfigure}
        
\end{sidewaysfigure}


\newpage
\clearpage
\begin{table}[ht]
\caption{\textbf{Descriptive Statistics}\label{table: descriptives}\\
This table presents the summary statistics of the main variables of interest for our sample. \textit{$\Delta$ Uncertainty} is the change in uncertainty expressions for the manuscript, from the pre-submission version to the published version. \textit{Female} is an indicator variable equal to one if the manuscript is authored by at least one female, and equal to zero if the manuscript is authored by an all-male team. \textit{Citations} are calculated as an equal weighting of each authors' Google Scholar citations (in thousands). \textit{Review Time} is the number of years between the pre-submission version of the manuscript and the publication year. \textit{$\Delta$ Uncertainty}, \textit{Female}, \textit{Citations}, and \textit{Review Time} are all measured at the paired (pre-submission paired with published) manuscript level. \textit{GGI} is the gender gap index, measured at the editor's country-of-origin level; the GGI is obtained from the World Economic Forum and ranges from 0 to 100, with higher values indicating greater parity. \textit{Male Editor} is an indicator variable equal to one if the editor is male, zero otherwise. \textit{\%Male Co-authorships} is calculated as the editor's average percent of male co-authors relative to total co-authors (omitting the editor his or herself in this calculation). \textit{Graduation Year} is the year that the editor graduated with a PhD. \textit{GGI}, \textit{Male Editor}, \textit{\%Male Co-authorships}, and \textit{Graduation Year} are all measured at the editor level.}
\centering\medskip
\setkeys{Gin}{keepaspectratio}
\centering\medskip
\scalebox{1}{
\begin{tabular}{lcccccc}
\hline\hline
				
		&\textit{Obs.}&\textit{Mean}&\textit{SD}&\textit{Q1}&\textit{Median}&\textit{Q3}\\\hline
$\Delta$ Uncertainty	&3,406	&3.354&	65.587&	-20&	3&	27\\
Female	&3,406&	0.343&	0.475	&0	&0	&1\\
Citations (Thousands)	&3,406	&7.057	&14.268	&1.163	&3.118	&7.168\\
Review Time&	3,406	&2.629	&1.801	&1	&2	&4\\
GGI&	88	&73.916	&4.884	&72.1	&76.3	&76.3\\

Male Editor&	88	&0.841	&0.368	&1	&1	&1\\
\%Male Co-authorships&	88	&83.833	&12.064	&76.941	&85.898	&93.685\\
Graduation Year&	88	&1994	&8.199	&1989	&1996	&2000\\
\hline \hline
	\end{tabular}}

\end{table}

\newpage
\begin{sidewaystable}[ht]
\caption{\textbf{Difference in Mean Values of Uncertainty Words}\label{table: genderdescriptives}\\
This table reports univariate differences in uncertainty words for female versus male authored papers. Papers are defined as female authored if at least one author is a female. Panel A reports gender differences in uncertainty for the pre-submission version of the manuscript, the published version of the manuscript, and the changes in uncertainty from pre-submission to published versions. Panel B reports similar statistics, but we split the dependent variable (uncertainty expressions) into its two components: hedge words are obtained through our SciBERT hedge detection model, and BOW are thirteen additional words included as a bank of words. Panel C reports changes in uncertainty expressions for female authors and male authors, split by the sections that uncertainty appears in the paper.}
\centering\medskip
 \setkeys{Gin}{keepaspectratio}
 \subcaption{Panel A: Gender Differences in Uncertainty}
\scalebox{1}{
\begin{tabular}{lcccc}
\hline \hline
& \textit{Obs.}&	\textit{Uncert.: Pre-Sub.}	&\textit{Uncert.: Pub}	&	\textit{$\Delta$Uncert.}\\\hline
Female&1,167 &207.630&214.776&7.147\\
Male&2,239&202.604&203.980&1.377\\\hline
Difference	&&5.026&10.796***&5.770**\\

\hline \hline
	\end{tabular}}

\subcaption{Panel B: Gender Differences -- Split by Hedge Words and Bank of Words}
\centering\medskip
\setkeys{Gin}{keepaspectratio}
\scalebox{1}{
\begin{tabular}{lccccccc}
\hline \hline
&&\multicolumn{3}{c}{Hedge Words}&\multicolumn{3}{c}{BOW}\\
\cmidrule(lr){3-5}
\cmidrule(lr){6-8}
&	\textit{Obs.} &\textit{Pre-Sub}	&\textit{Published}	&	\textit{$\Delta$ }&	\textit{Pre-Sub}	&\textit{Published}	&	\textit{$\Delta$ }\\\hline

Female	&1,167&188.662&195.042&6.380&18.967&19.734&0.767\\
Male&2,239&183.839&185.039&1.200&18.764&18.941&0.177\\\hline
Difference	&&4.823&10.003***&5.180**&0.203&0.793*&0.590*\\

\hline \hline
	\end{tabular}}
	
\subcaption{Panel C: Gender differences by section}
\centering\medskip
\setkeys{Gin}{keepaspectratio}
\scalebox{1}{
\begin{tabular}{lccccc}
\hline \hline
&&\multicolumn{4}{c}{$\Delta$ Uncertainty}\\\hline
&\textit{Obs.}&						\textit{Abstract}	&\textit{Introduction}	&	\textit{Conclusion}&	\textit{Full Text}\\\hline
Female&1,167&-0.045	&-1.598&0.829&7.147\\
Male&2,239&-0.042&-2.573&-0.138&1.377\\\hline
Difference&	&-0.003&0.974*&0.967**&5.770**\\

\hline \hline
	\end{tabular}}

\end{sidewaystable}

\newpage
\begin{table}[ht]
\caption{\textbf{Mean Uncertainty Change by Field and by Journal}\label{table: byjournal}\\
This table reports univariate differences in uncertainty words for female versus male authored papers. Papers are defined as female authored if at least one author is a female. We report statistics separately by the field in which the paper was published and by the journal in which it is published.}
\centering\medskip
 \setkeys{Gin}{keepaspectratio}
\scalebox{1}{
\begin{tabular}{lcccccc}
\hline \hline
&\multicolumn{2}{c}{\textit{Female}}&\multicolumn{2}{c}{\textit{Male}}\\
\cmidrule(lr){2-3}
\cmidrule(lr){4-5}
&\textit{Obs.}&\textit{$\Delta$Uncert.}&\textit{Obs.}&\textit{$\Delta$Uncert.}&Diff.\\\hline
\textit{Economics}&347&0.578&820&-5.829&6.406\\\hline
&&&&&\\
\hspace{4mm}American Economic Review&146&-10.568&270&-12.163&1.594\\
            &   &      & &&  \\
\hspace{4mm}Econometrica&42&-4.571&153&-17.876&13.304\\
   &      &  & &&   \\
\hspace{4mm}Journal of Political Economy&45&0.444&125&-18.120&18.564\\
   &      &      &&&\\  
\hspace{4mm}Quarterly Journal of Economics&56&0.848&109&19.284&-18.436\\
&   &      & &&     \\
\hspace{4mm}Review of Economic Studies&58&32.207&163&8.604&23.603*\\
&   &      &  &&    \\\hline

\textit{Finance}&328&17.291&713&7.903&9.389**\\\hline
&&&&&\\
\hspace{4mm}Journal of Finance&74&22.959&172&4.494&18.465**\\
   &      &  &   && \\
\hspace{4mm}Journal of Financial Economics&147&15.935&282&10.069&5.866\\
   &      &      &&&\\
\hspace{4mm}Review of Financial Studies&107&15.234&259&7.807&7.427\\
&   &      &    &&  \\\hline

\textit{Accounting}&172&12.378&216&0.722&11.655*\\\hline
&&&&&\\
\hspace{4mm}The Accounting Review&72&5.750&73&-7.712&13.462*    \\
            &      &      &&&\\
\hspace{4mm}Journal of Accounting and Economics&59&22.898&88&-1.511&24.410**\\
   &      &    & && \\
\hspace{4mm}Journal of Accounting Research&41&8.878&55&15.491&-6.613\\
&   &      &   &&   \\\hline        

\textit{Management}&320&1.059&490&4.228&-3.168\\\hline
&&&&&\\            
\hspace{4mm}Academy of Management Journal&32&8.469&17&2.176&6.292\\
&&&&&\\
\hspace{4mm}Academy of Management Review&8&-32.5&12&3.167&-35.667\\
&&&&&\\

\hspace{4mm}Administrative Science Quarterly&23&-4.957&16&-14.750&9.793\\
&&&&&\\
\hspace{4mm}Management Science&257&1.720&445&5.017&-3.297\\
   &&&&&\\

            \hline \hline
	\end{tabular}}
	
\end{table}

\newpage
\begin{sidewaystable}[ht]
\caption{\textbf{Multivariate regression estimating changes in uncertainty}\label{table: primary}\\
This table reports the results of estimating equation \ref{eq: main}, where observations are at the paired (pre-submission to published) manuscript level, and the dependent variable is the change in uncertainty expressions from the pre-submission to the published version of the manuscript. All other variables are defined previously. Column 1 estimates the results for the full text, and columns 2, 3, and 4 estimates the changes in uncertainty occurring in the abstract, the introduction, and the conclusion, respectively. In column 5, we again estimate the changes in uncertainty in the full text document, but we add the change in \textit{total words} from pre-submission to published version as an additional control variable. In column 6, we log transform the dependent variable and re-estimate the results. Standard errors are clustered at the topic level and are reported in parentheses. *** significant at 1\%, ** significant at 5\%, * significant at 10\%.}
 \setkeys{Gin}{keepaspectratio}
\centering\medskip

 \scalebox{1}{
\begin{tabular}{lcccccc}
\hline \hline
&\multicolumn{5}{c}{$\Delta$ Uncertainty}&\multicolumn{1}{c}{Log($\Delta$Uncertainty)}\\\hline

	  &\multicolumn{1}{c}{(1)}         &\multicolumn{1}{c}{(2)}         &\multicolumn{1}{c}{(3)}         &\multicolumn{1}{c}{(4)}&\multicolumn{1}{c}{(5)}         &\multicolumn{1}{c}{(6)} \\\hline
	  &\multicolumn{1}{c}{Fulltext}         &\multicolumn{1}{c}{Abstract}         &\multicolumn{1}{c}{Introduction}         &\multicolumn{1}{c}{Conclusion}&\multicolumn{1}{c}{Fulltext}         &\multicolumn{1}{c}{Fulltext} \\\hline
           
Female  &        5.276***&       0.019   &       0.757   &       0.899*  &       2.399*  &       0.120** \\
            &     (1.863)   &     (0.088)   &     (0.568)   &     (0.518)   &     (1.344)   &     (0.052)   \\
            
Citation Index &      0.083   &      -0.001   &       0.020   &       0.034** &      -0.112***&      -0.003   \\
            &     (0.054)   &     (0.003)   &     (0.017)   &     (0.015)   &     (0.037)   &     (0.003)   \\
Review Time    &       4.081** &      -0.053*  &       0.382*  &       0.512***&      -0.508   &       0.035   \\
            &     (1.642)   &     (0.029)   &     (0.197)   &     (0.139)   &     (0.784)   &     (0.027)   \\
$\Delta$ Total Words&               &               &               &               &       0.015***&       0.001***\\
            &               &               &               &               &     (0.001)   &     (0.000)       \\\hline
Observations    &   3,406   &        3,406   &        3,406   &        3,406   &        3,406   &        3,406   \\
Adjusted \(R^{2}\)&       0.031   &       0.043   &       0.068   &       0.026   &       0.661   &       0.539   \\
JournalFE   &        Y       &     Y          &      Y         &       Y        &        Y       &      Y             \\
YearFE      &         Y      &       Y        &       Y        &        Y       &       Y        &      Y              \\
TopicFE     &          Y     &       Y        &        Y       &       Y        &       Y        &       Y                  \\
\hline \hline
	\end{tabular}}
\end{sidewaystable}

\newpage
\begin{table}[ht]
\caption{\textbf{Robustness: NLP model to predict counterfactual changes in hedge expressions for `pseudo female' observations}\label{table: pseudo}\\
This table reports the results of our NLP model for causal analysis. Specifically, we build a model that takes as inputs all fundamentals, including the text with deleted uncertainty words, and non-textual meta information. We then train the model using two fully-connected neural networks to predict the number of uncertainty expressions. Finally, we extract all articles written by male authors, feed them into our NLP model and artificially flip the gender from male to female to predict a counterfactual, all-else-equal, change in uncertainty for a pseudo-female observation. Details of the model architecture are outlined in Figure \ref{fig:concat_bert_arch}.}
\centering\medskip
 \setkeys{Gin}{keepaspectratio}
\scalebox{1}{
\begin{tabular}{lc}
\hline \hline
			& 		\textit{$\Delta$ Uncertainty Words}\\\hline
True Male&201.79\\
Counterfactual ``Pseudo'' Female	&207.53\\\hline
Difference	&5.74***\\\hline
Model Fit&0.8080\\

\hline \hline
	\end{tabular}}
	\end{table}

\newpage
\begin{sidewaystable}[ht]
\caption{\textbf{Editor's Impact on Female Changes in Uncertainty}\label{table: editorfe}\\
In panel A, we report the fit of a benchmark specification estimating the relationship between changes in uncertainty words and a list of control variables and fixed effects. The benchmark specification includes controls for the number of years the manuscript was in the review process, the change in the total number of words, and fixed effects for the journal, topic, year published, and author. Regressions are estimated at the individual author-article level, which allows us to include the author fixed effect (e.g., if an article has three authors, we include three observations: one article for each author). In the second row, we add an interactive fixed effect -- editor*female-- and report the adjusted R-square and the F-test for joint significance of the editor*female fixed effects. In the third and fourth rows, we estimate the F-tests for only those editor-fixed effects where the editor is randomly assigned manuscripts (Journal of Accounting and Economics, and Journal of Accounting Research, respectively). In panel B, we report the size distribution of the editor*Female fixed effect.}
\centering\medskip
 \setkeys{Gin}{keepaspectratio}
 \subcaption{Panel A: Testing the joint significance of editor*female fixed effects.}
\scalebox{1}{
\begin{tabular}{lccc}
\hline \hline
&\multicolumn{3}{c}{$\Delta$ Uncertainty Words}\\\hline
&\multicolumn{1}{c}{P-value}&\multicolumn{1}{c}{N}&\multicolumn{1}{c}{Adjusted \(R^{2}\)}\\

\textit{Author*Article-level Regressions}&&&\\\hline
\hspace{4mm}Journal, Topic, Year, and Author FE&&2,864&0.678\\
\hspace{4mm}Journal, Topic, Year, Author, and Editor*Female FE&15.04(<0.001)&2,864&0.706\\
\hspace{40mm}\textit{Randomized editors: JAE} &2.05 (0.056)&&\\
\hspace{40mm}\textit{Randomized editors: JAR} &3.36 (0.010)&&\\\hline
\hline \hline
	\end{tabular}}

\subcaption{Panel B: Size distribution of editor*female fixed effects}
\centering\medskip
 \setkeys{Gin}{keepaspectratio}

\scalebox{1}{
\begin{tabular}{lccccc}
\hline \hline
&\multicolumn{1}{c}{N}&\multicolumn{1}{c}{SD}&\multicolumn{1}{c}{P25}&\multicolumn{1}{c}{P50}&\multicolumn{1}{c}{P75}\\\hline	
Editor*Female FE	&88&50.211&-3.507&14.755&49.918\\
\hline\hline
	\end{tabular}}
	\end{sidewaystable}

	\newpage

\begin{sidewaystable}[ht]
\caption{\textbf{Correlation between the editor*female fixed effect and characteristics of the editor}\label{table: corrmain}\\
This table reports the correlation between the editor's author-gender gap and characteristics of the editor. Each editor's author-gender gap is obtained by extracting the editor*female fixed effect, $\gamma_{j}$, from estimating equation \ref{eq: second}. We weight each fixed effect by the inverse of the square of the standard error to account for estimation error.\textit{GGI} is the gender gap index, measured at the editor's country-of-origin level; the GGI is obtained from the World Economic Forum and ranges from 0 to 100, with higher values indicating greater parity. \textit{Male Editor} is an indicator variable equal to one if the editor is male, zero otherwise. \textit{\%Male Co-authorships} is calculated as the editor's average percent of male co-authors relative to total co-authors (omitting the editor his or herself in this calculation). \textit{Graduation Year} is the year that the editor graduated with a PhD.}
\centering\medskip
 \setkeys{Gin}{keepaspectratio}

\scalebox{1}{
\begin{tabular}{lccccc}
\hline \hline
&\multicolumn{1}{c}{Editor*Female FE}&\multicolumn{1}{c}{GGI}&\multicolumn{1}{c}{Male Editor}&\multicolumn{1}{c}{Male Coauthorships}&\multicolumn{1}{c}{Graduation Year}\\\hline
Editor*Female FE	&1.000		&		&&&\\
&&&&&\\
GGI	&-0.235**&	1.000&&&	\\		
	&(0.027)	&&&&\\			
					
Male Editor	&0.055&	0.285**&	1.000&&\\		
	&(0.614)	&(0.007)	&&&\\		
					
\%Male Co-authorships&	0.320**&	-0.195*&0.217**&	1.000&\\	
	&(0.002)&(0.069)&	(0.043)&&\\		
					
Graduation Year&	-0.227**&	-0.197*&	0.051&	0.048&	1.000\\
	&(0.032)&	(0.064)&	(0.637)&	(0.657)&\\
	\hline\hline	

	\end{tabular}}
	\end{sidewaystable}

\begin{table}[ht]
\caption{\textbf{Regression results of relation between the editor*female fixed effect and characteristics of the editor.}\label{table: editorFEmain}\\
This table reports the results of regressing the editor's author-gender gap on his or her characteristics. Each editor's author-gender gap is obtained by extracting the editor*female fixed effect, $\gamma_{j}$, from estimating equation \ref{eq: second}. We weight each fixed effect by the inverse of the square of the standard error to account for estimation error.\textit{GGI} is the gender gap index, measured at the editor's country-of-origin level; the GGI is obtained from the World Economic Forum and ranges from 0 to 100, with higher values indicating greater parity. \textit{Male Editor} is an indicator variable equal to one if the editor is male, zero otherwise. \textit{\%Male Co-authorships} is calculated as the editor's average percent of male co-authors relative to total co-authors (omitting the editor his or herself in this calculation). \textit{Graduation Year} is the year that the editor graduated with a PhD. *** significant at 1\%, ** significant at 5\%, * significant at 10\%.}
\centering\medskip
 \setkeys{Gin}{keepaspectratio}

\scalebox{1}{
\begin{tabular}{lc}
\hline \hline
&\multicolumn{1}{c}{Editor*Female Fixed Effect}\\\hline
GGI  &       -2.522**   \\
                    &     (1.042)     \\

Male Editor    &     7.795     \\
                  &    (15.082)    \\
\%Male Co-authorships &     1.281**   \\
            &     (0.488)       \\
Graduation Year     &      -2.057**  \\
                   &     (0.695)  \\\hline

Observations        &           88     \\
Adj. \(R^{2}\)          &       0.181     \\
\hline\hline

	\end{tabular}}
	\end{table}

\newpage
\clearpage
\begin{sidewaystable}[ht]
\caption{\textbf{Robustness Analyses}\label{table: robust}\\
This table presents robustness tests to assess whether the author-gender gap arises due to (1) higher gender-neutral strictness of certain editors, or (2) because certain editors are more lenient on females in the accept-reject decision. Our first proxy for editor strictness include the Editor*Male fixed effect, which is estimated as the main effect on the editor index, estimated from equation \ref{eq: second}; the second proxy for editor strictness is his or her overall acceptance rate, calculated as the number of articles the editor accepts relative to the total number of articles published in that particular journal. Finally, we assess the editor's leniency on women authors through the proxy \textit{Female Acceptance Rate}, which is calculated as the total number of women-authored papers accepted by the editor, relative to the total number of papers that editor accepts.}
\centering\medskip
 \setkeys{Gin}{keepaspectratio}

\scalebox{1}{
\begin{tabular}{lcccc}
\hline \hline
&\multicolumn{1}{c}{Editor*Female FE}&\multicolumn{1}{c}{Editor*Male FE}&\multicolumn{1}{c}{Acceptance Rate}&\multicolumn{1}{c}{Female Acceptance Rate}\\\hline
Editor*Female FE  &	1.000	&	&	&\\
             &&&&\\
Editor*Male FE&	0.130&	1.000&	&\\
&	(0.222)&	&&\\
Acceptance Rate&	-0.146&	0.050	&1.000&\\
&	(0.169)&	(0.637)&&\\
Female Acceptance Rate&	0.051	&-0.240**&	-0.491***	&1.000\\
&	(0.630)&	(0.022)&	(0.000)&\\
\hline\hline

	\end{tabular}}
	\end{sidewaystable}

\setcounter{table}{0}
\setcounter{figure}{0}
\newpage
\clearpage
\section{Appendix}
\renewcommand{\thetable}{A\arabic{table}}
\renewcommand{\thefigure}{A\arabic{figure}}

\newpage
\clearpage
\textbf{A1. LDA Topics}

Below is a list of the topic categories, as produced by LDA. For each topic, we list the keywords and their relative weighting in each topic category.
\label{table:LDA}

\textbf{Topic: 0} 
\begin{itemize}
\item{Words: 0.024*"treatment" + 0.017*"experiment" + 0.016*"participant" + 0.009*"task" + 0.008*"bias" + 0.007*"belief" + 0.007*"experimental" + 0.006*"effort" + 0.006*"survey" + 0.006*"preference"}
\end{itemize}

\textbf{Topic: 1}
\begin{itemize} 
\item{Words: 0.024*"board" + 0.021*"director" + 0.019*"political" + 0.015*"vote" + 0.012*"election" + 0.011*"member" + 0.011*"party" + 0.010*"voting" + 0.010*"local" + 0.010*"committee"}
\end{itemize}

\textbf{Topic: 2} 
\begin{itemize}
\item{Words: 0.051*"worker" + 0.050*"wage" + 0.033*"labor" + 0.024*"job" + 0.020*"employment" + 0.012*"unemployment" + 0.012*"employee" + 0.010*"productivity" + 0.008*"earnings" + 0.008*"hour"}
\end{itemize}

\textbf{Topic: 3} 
\begin{itemize}
\item{Words: 0.060*"analyst" + 0.058*"forecast" + 0.042*"earnings" + 0.016*"news" + 0.012*"announcement" + 0.011*"coverage" + 0.009*"enforcement" + 0.009*"quarter" + 0.009*"revision" + 0.009*"medium"}
\end{itemize}

\textbf{Topic: 4} 
\begin{itemize}
\item{Words: 0.020*"quality" + 0.019*"city" + 0.012*"jt" + 0.012*"insurance" + 0.012*"plan" + 0.011*"advertising" + 0.011*"percent" + 0.009*"health" + 0.009*"payment" + 0.008*"consumer"}
\end{itemize}

\textbf{Topic: 5}
\begin{itemize} 
\item{Words: 0.072*"loan" + 0.033*"credit" + 0.029*"bank" + 0.028*"borrower" + 0.017*"lender" + 0.015*"mortgage" + 0.014*"lending" + 0.011*"covenant" + 0.009*"house" + 0.009*"default"}
\end{itemize}

\textbf{Topic: 6}
\begin{itemize} 
\item{Words: 0.050*"consumer" + 0.035*"customer" + 0.020*"sale" + 0.020*"supplier" + 0.014*"search" + 0.013*"purchase" + 0.012*"retailer" + 0.011*"inventory" + 0.010*"user" + 0.010*"store"}
\end{itemize}

\textbf{Topic: 7}
\begin{itemize} 
\item{Words: 0.023*"volatility" + 0.018*"portfolio" + 0.014*"premium" + 0.012*"variance" + 0.012*"option" + 0.011*"investor" + 0.009*"pricing" + 0.008*"equity" + 0.007*"consumption" + 0.007*"excess"}
\end{itemize}

\textbf{Topic: 8} 
\begin{itemize}
\item{Words: 0.035*"shock" + 0.018*"growth" + 0.014*"consumption" + 0.012*"economy" + 0.012*"output" + 0.012*"inflation" + 0.008*"dynamic" + 0.008*"monetary" + 0.007*"household" + 0.006*"productivity"}
\end{itemize}

\textbf{Topic: 9}
\begin{itemize} 
\item{Words: 0.039*"audit" + 0.036*"auditor" + 0.021*"accounting" + 0.020*"client" + 0.018*"ipo" + 0.013*"quality" + 0.011*"company" + 0.010*"statement" + 0.008*"opinion" + 0.007*"internal"}
\end{itemize}

\textbf{Topic: 10}
\begin{itemize} 
\item{Words: 0.038*"ceo" + 0.021*"target" + 0.019*"shareholder" + 0.014*"ownership" + 0.013*"acquisition" + 0.012*"deal" + 0.012*"merger" + 0.011*"compensation" + 0.011*"option" + 0.010*"executive"}
\end{itemize}

\textbf{Topic: 11}
\begin{itemize} 
\item{Words: 0.010*"agent" + 0.010*"let" + 0.009*"utility" + 0.009*"optimal" + 0.009*"theorem" + 0.007*"preference" + 0.006*"solution" + 0.006*"proof" + 0.005*"vector" + 0.005*"random"}
\end{itemize}

\textbf{Topic: 12}
\begin{itemize} 
\item{Words: 0.012*"organization" + 0.009*"social" + 0.007*"organizational" + 0.005*"member" + 0.004*"resource" + 0.004*"employee" + 0.004*"identity" + 0.004*"people" + 0.004*"task" + 0.003*"action"}
\end{itemize}

\textbf{Topic: 13}
\begin{itemize} 
\item{Words: 0.024*"patent" + 0.022*"innovation" + 0.018*"technology" + 0.013*"entry" + 0.010*"knowledge" + 0.009*"competition" + 0.006*"technological" + 0.006*"incumbent" + 0.006*"entrant" + 0.006*"alliance"}
\end{itemize}

\textbf{Topic: 14}
\begin{itemize} 
\item{Words: 0.048*"fund" + 0.030*"investor" + 0.028*"trading" + 0.020*"trade" + 0.014*"day" + 0.010*"portfolio" + 0.010*"liquidity" + 0.009*"trader" + 0.009*"volume" + 0.006*"announcement"}
\end{itemize}

\textbf{Topic: 15} 
\begin{itemize}
\item{Words: 0.030*"network" + 0.021*"patient" + 0.017*"auction" + 0.016*"bid" + 0.014*"hospital" + 0.012*"bidder" + 0.010*"player" + 0.008*"service" + 0.005*"delay" + 0.005*"pollution"}
\end{itemize}

\textbf{Topic: 16}
\begin{itemize} 
\item{Words: 0.027*"earnings" + 0.021*"cash" + 0.014*"portfolio" + 0.013*"flow" + 0.011*"cash\_flow" + 0.011*"month" + 0.009*"accrual" + 0.008*"equity" + 0.007*"relation" + 0.007*"growth"}
\end{itemize}

\textbf{Topic: 17} 
\begin{itemize}
\item{Words: 0.036*"student" + 0.032*"school" + 0.020*"woman" + 0.013*"program" + 0.012*"gender" + 0.012*"score" + 0.012*"education" + 0.011*"female" + 0.011*"men" + 0.011*"college"}
\end{itemize}

\textbf{Topic: 18} 
\begin{itemize}
\item{Words: 0.094*"tax" + 0.044*"income" + 0.014*"government" + 0.010*"dividend" + 0.008*"cash" + 0.007*"corporate" + 0.007*"percent" + 0.007*"saving" + 0.007*"foreign" + 0.006*"revenue"}
\end{itemize}

\textbf{Topic: 19} 
\begin{itemize}
\item{Words: 0.044*"country" + 0.018*"trade" + 0.014*"production" + 0.011*"foreign" + 0.008*"sector" + 0.007*"export" + 0.007*"domestic" + 0.007*"input" + 0.006*"international" + 0.006*"exchange"}
\end{itemize}

\textbf{Topic: 20}
\begin{itemize} 
\item{Words: 0.056*"bank" + 0.032*"debt" + 0.024*"bond" + 0.016*"rating" + 0.014*"default" + 0.013*"credit" + 0.010*"equity" + 0.009*"liquidity" + 0.009*"leverage" + 0.008*"crisis"}
\end{itemize}

\textbf{Topic: 21}
\begin{itemize} 
\item{Words: 0.026*"household" + 0.018*"income" + 0.013*"age" + 0.013*"child" + 0.009*"consumption" + 0.009*"family" + 0.009*"wealth" + 0.008*"population" + 0.008*"percent" + 0.007*"county"}
\end{itemize}

\textbf{Topic: 22} 
\begin{itemize}
\item{Words: 0.035*"equilibrium" + 0.014*"agent" + 0.013*"contract" + 0.012*"proposition" + 0.012*"optimal" + 0.011*"buyer" + 0.011*"seller" + 0.009*"signal" + 0.009*"profit" + 0.009*"payoff"}
\end{itemize}

\textbf{Topic: 23} 
\begin{itemize}
\item{Words: 0.037*"disclosure" + 0.016*"accounting" + 0.012*"reporting" + 0.008*"investor" + 0.007*"indicator" + 0.007*"fair" + 0.007*"adoption" + 0.006*"sec" + 0.006*"filing" + 0.006*"percent"}
\end{itemize}

\textbf{Topic: 24} 
\begin{itemize}
\item{Words: 0.130*"manager" + 0.017*"team" + 0.016*"compensation" + 0.014*"tie" + 0.013*"venture" + 0.013*"managerial" + 0.010*"founder" + 0.008*"manipulation" + 0.008*"division" + 0.008*"partner"}
\end{itemize}

\newpage
\textbf{A2. Sample Editor Letter}

Below is a sample of an editor letter, with sensitive identifying information redacted. The letter was provided to the authors in the third ``round'' of revisions.
\label{a2:editorletter}
\includepdf[pages={1-},scale=0.9]{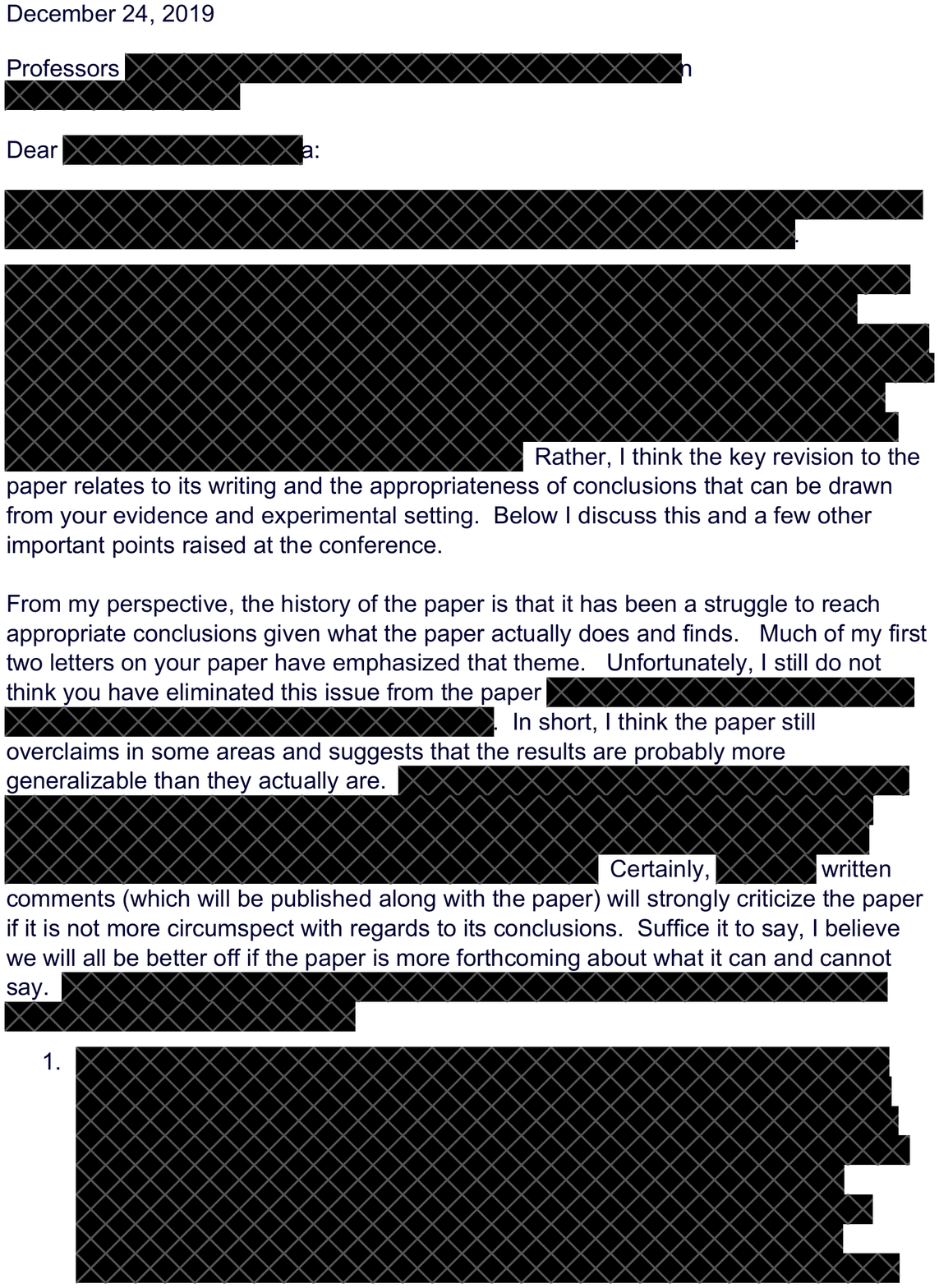}

\newpage
\textbf{A3. Email Responses from Journal Managers}

Below are two email exchanges between the authors and journal managers to inquire about manuscript assignment policies. Both indicate that assignment is random. 
\label{a3:journalletter1}


\section*{Supplementary material}

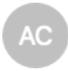

**From:** Fedorova, Ekaterina <<u>Ekaterina.Fedorova@chicagobooth.edu</u>>
**Sent:** Wednesday, March 30, 2022 5:10 PM
**To:** Lisa M. Heiberger <<u>Lisa.heiberger@chicagobooth.edu</u>>
**Subject:** Inquiry

Dear Lisa Heiberger,
My name is Ekaterina Fedorova, and I am a research professional for Professor Anna Costello at the University of Chicago. We are working on a project related to the publication process, and we would like to understand how submissions are handled at the Journal of Accounting Review. Specifically, we'd like to know how manuscripts are assigned to editors.

Can you confirm whether submitted manuscripts are assigned to editors:
a) Through random assignment (subject to editor availability and conflicts of interest)
b) According to topic/expertise
or
c) Other (please specify)
Thank you in advance for your response.
Best regards,
Ekaterina Fedorova

Begin forwarded message:

**From:** "Lisa M. Heiberger" <<u>Lisa.heiberger@chicagobooth.edu</u>>
**Subject: RE: Inquiry**
**Date:** April 7, 2022 at 7:16:05 AM CDT
**To:** "Fedorova, Ekaterina" <<u>Ekaterina.Fedorova@chicagobooth.edu</u>>
**Cc:** "Costello, Anna" <<u>Anna.Costello@chicagobooth.edu</u>>, "Lisa M. Heiberger" <<u>Lisa.heiberger@chicagobooth.edu</u>>

Dear Ekaterina,

JAR papers are assigned

a) Through random assignment (subject to editor availability and conflicts of interest)

Best,
Lisa
**Lisa M. Heiberger**
Editorial Manager, Journal of Accounting Research
**Chicago Booth**

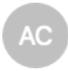

**From:** Fedorova, Ekaterina <Ekaterina.Fedorova@chicagobooth.edu>
**Sent:** Wednesday, March 30, 2022 6:10 PM
**To:** Fardone, Frank <fardone@wharton.upenn.edu>
**Subject:** Inquiry

Dear Frank Fardone,

My name is Ekaterina Fedorova, and I am a research professional for Professor Anna Costello at the University of Chicago. We are working on a project related to the publication process, and we would like to understand how submissions are handled at the Journal of Accounting and Economics. Specifically, we'd like to know how manuscripts are assigned to editors.

Can you confirm whether submitted manuscripts are assigned to editors:
a) Through random assignment (subject to editor availability and conflicts of interest)
b) According to topic/expertise
or
c) Other (please specify)

Thank you in advance for your response.

Best regards,

Ekaterina Fedorova

**From:** "Fardone, Frank" <fardone@wharton.upenn.edu>
**Date:** Thursday, March 31, 2022 at 8:02 AM
**To:** "Fedorova, Ekaterina" <Ekaterina.Fedorova@chicagobooth.edu>
**Subject:** RE: Inquiry

Dear Ms. Fedorova,

The JAE assigned editors to new manuscripts through random assignment (subject to editor availability and conflicts of interest).

Sincerely,
Frank Fardone
Editorial Assistant
*Journal of Accounting & Economics*
Accounting Department
The Wharton School
University of Pennsylvania
Philadelphia, PA 19104
215-898-4942



\end{document}